\pdfoutput=1
\documentclass[aps,prx,twocolumn,longbibliography,superscriptaddress]{revtex4-1}
\usepackage{amsmath}
\usepackage{bm}
\usepackage{bbold}
\usepackage{amssymb}
\usepackage{graphicx}
\usepackage[ hidelinks = true, colorlinks = true, citecolor = blue, urlcolor = blue, linkcolor = blue]{hyperref}
\usepackage{verbatim}
\usepackage{color}
\usepackage[capitalise]{cleveref}
\usepackage[normalem]{ulem}

\allowdisplaybreaks[3]

\begin{document}

\title{Coherent backaction between spins and an electronic bath: {N}on-{M}arkovian dynamics and low-temperature quantum thermodynamic electron cooling}

\author{Stephanie Matern}
\affiliation{SUPA, School of Physics and Astronomy, University of St Andrews, North Haugh, St Andrews KY16 9SS, United Kingdom}
\author{Daniel Loss}
\affiliation{Department of Physics, University of Basel, Klingelbergstrasse 82, CH-4056 Basel, Switzerland}
\author{Jelena Klinovaja}
\affiliation{Department of Physics, University of Basel, Klingelbergstrasse 82, CH-4056 Basel, Switzerland}
\author{Bernd Braunecker}
\affiliation{SUPA, School of Physics and Astronomy, University of St Andrews, North Haugh, St Andrews KY16 9SS, United Kingdom}

\date{\today}

\begin{abstract}
We provide a versatile analytical framework for calculating the dynamics of a spin system in contact with a fermionic bath beyond the Markov approximation.
The approach is based on a second order expansion of the Nakajima-Zwanzig master equation but systematically includes all quantum coherent memory effects leading to non-Markovian dynamics.
Our results describe, for the free induction decay, the full time range from the non-Markovian dynamics at short times, to the well-known exponential thermal decay at long times. We provide full analytic results for the entire time range using a bath of itinerant electrons as an archetype for universal quantum fluctuations. Furthermore, we propose a quantum thermodynamic scheme to employ the temperature insensitivity of the non-Markovian decay to transport heat out of the electron system and thus, by repeated reinitialization of a cluster of spins, to efficiently cool the electrons at very low temperatures.
\end{abstract}

\maketitle


%

\section{Introduction}

\begin{figure}[t]
    \centering
    \includegraphics[width=0.9\columnwidth]{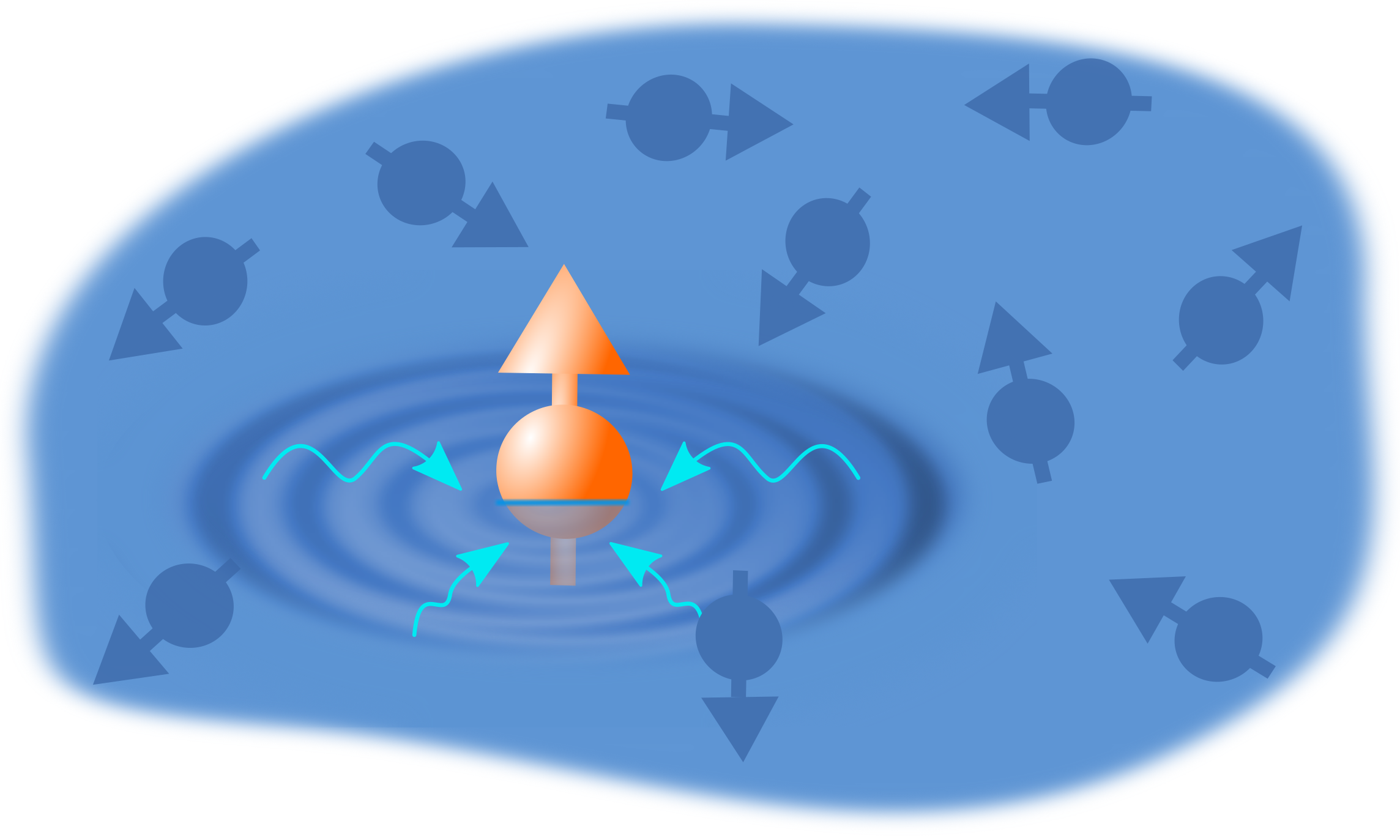}
    \caption{
	Sketch of the type of system under consideration. A localised spin $I$ (orange arrow) is embedded in a fermionic conductor. The interaction with the itinerant spins (collective of dark arrows) creates a magnetic excitation in the conductor (ripples) that shapes the dynamics of $I$ through backaction (bright wiggled arrows). At short enough times the magnetic excitation remains quantum coherent, causing a non-Markovian imprint on the evolution of $I$.
    \label{fig:system}}
\end{figure}

In recent years, quantum effects have become controllable, to the extent that a proper quantum technology based on the core properties of quantum parallelism and
entanglement is on the rise. The success of this technology will rely on our ability to understand
and control more intricate quantum states and their evolution. One of the main challenges is to control the interaction of a quantum system with its environment \cite{Kloeffel,Yang,Sun}. Such an interaction is detrimental for a quantum application as it leads to the major issue
of decoherence,
but it also can be used positively, as it is on the basis of quantum thermodynamics \cite{Kosloff,Goold,Vinjanampathy}, because it can be employed
to manipulate the system via the environment \cite{Stace,Haertle} and to reveal information about the environment itself.
The latter is, for instance, at the core of magnetic resonance techniques such as nuclear magnetic resonance and electron paramagnetic Resonance in which the measurable decay of individual quantum spins is the result of their interaction with their environment \cite{Ihara,Mamin2007,Augustyniak,Hanson}. Figure \ref{fig:system} shows a sketch of such a system.

It is the goal of this paper to provide an analytical framework to access the system-environment correlations, and thus to provide a direct access to the core physics that connects the broad range of physical disciplines from the mature field of magnetic resonance to the recent development of quantum thermodynamics. Although we restrict ourselves to the weak coupling limit, we provide a systematic approach to include all Markovian and non-Markovian dynamics. This allows us to bridge fully from the quantum coherent regime at short time scales to the thermodynamic regime at long time scales.

In common situations the number of degrees of freedom of the environment is macroscopic. Then, although the overall evolution remains unitary, the environment acts as a bath in which any transmitted information is effectively dispersed irreversibly. The environment is called Markovian or memoryless if during the contact with the system it remains unaffected by the interaction in its equilibrium state.
Since a zero memory time cannot exist physically, the Markovian property is an approximation that is valid if the memory time is shorter than the characteristic time scales of the dynamics of the quantum system. While this approximation sufficiently describes quantum thermodynamics, decoherence or the behavior of magnetic resonance and similar techniques in many situations, it misses that during the memory time, even if it is short, the system and environment have a joint evolution in which in particular the excitation of the environment can act back on the system. This partially coherent backaction not only shapes the system's short time dynamics but can also leave an imprint in the long time behavior in the form of a correction to the expected Markovian dynamics \cite{Hell,DiVincenzo,Loss1998} or a modification of the performance in thermal machines \cite{Watanabe}. Such properties can thus be used passively as a diagnostic tool for the structure of the environment. But such non-Markovian behavior can also be used as a novel route to actively manipulate the environment through the quantum system \cite{Man} or to enhance the efficiency of quantum heat engines \cite{Abiuso}.

The non-Markovian treatment becomes pertinent when memory times are enhanced, such as the modern developments of reaching sub-millikelvin temperatures even for electronic nanostructures \cite{Clark,Palma1,Palma2}
and by the design and discovery of strongly correlated materials. The latter exhibit collective responses
to local excitations that naturally extend correlations in space and time, and thus the memory time.
Within the Markov approximation, only the spatial signatures of strong correlations have an impact
on the system's dynamics, which is a property routinely investigated for signatures of the strong correlation
physics \cite{Koelbl,Tuereci,Latta}. This neglects that the similarly strong temporal correlations delay the decay of memory
and that the dynamics must be complemented by a concise modeling of the non-Markovian dynamics of, for instance,
a spin's free decay in magnetic resonance.

When evolving a system under a Markovian master equation the bath is treated as a large memoryless reservoir. The state of a system in contact with such a reservoir thermalizes and decays exponentially fast. In the non-Markovian regime, the backaction onto the system by fluctuations in the bath is important as the times are too short for the memory kernel to decay sufficiently.

As an exact solution can rarely be found, different approaches exist to incorporate the short time dynamics and non-Markovian
effects based, for instance, on time-convolutionless master equations \cite{Timm,Fischer,Jing,Maniscalco} or the Nakajima-Zwanzig equation \cite{Nakajima,Zwanzig}. The Nakajima-Zwanzig equation is often used as starting point to set up hierarchical equation of motions \cite{Tanimura,Yan} or to modified projection methods \cite{DiVincenzo,Hartmann} including nonperturbative expansions \cite{LossCoish,Moustos} and re-summation techniques \cite{Reichman}.
Numerical algorithms include transfer tensors or path integral methods to reconstruct the memory kernel of the Nakajima-Zwanzig equation, which are even feasible beyond the weak coupling limit \cite{Pollock,Strathearn,Cao,Geva}.
Another possibility is to calculate the `initial slip' of the system and then evolving the state within a Markovian description with modified initial conditions \cite{Geigenmueller,Haake1985,Suarez,Yu}.
It should just be emphasised that any chosen approach has to maintain the positivity of the density matrix \cite{Loss1998,Whitney,Vacchini,Shabani}.
Closest in spirit to our approach is perhaps a Liouville-space decomposition method \cite{Hell, Hell_2014} with its emphasis on the importance of coherent backaction.

In this paper, we provide a theoretical framework for the entire cross-over from short to long times
for a spin system coupled to a fermionic system, starting from the exact Nakajima-Zwanzig equation.
We pursue three goals: (\emph{i}) the development of the general
formalism applicable to any spin and any environment system. (\emph{ii}) the analysis of non-Markovian behavior when
the environment is a Fermi gas, and (\emph{iii}) the demonstration that non-Markovianity can be used to actively
manipulate the environment through the proposition of a temperature independent quantum demagnetisation
cooling protocol that goes beyond a standard thermodynamic cycle.

For goal (\emph{i}), we set up a systematic analysis of the poles representing the various decay modes from the equation of
motion of the density matrix in Laplace space. The important feature is that we identify the ensemble of decay modes but perform approximations to the exact pole positions. This means that although we will use a second order expansion of the memory kernel we pick up the entity of decay modes and can cover with them the full time dynamics through all times. This approach is versatile and independent of the type of the environment, but to give a concrete example we focus on fermionic environments for the further study of non-Markovian behavior.
Although we expect the largest impact to arise from strongly correlated systems, we have chosen for goal (\emph{ii})
an environment in the form of a noninteracting itinerant fermionic system, sketched in \cref{fig:system}. This environment is as close to a memoryless fermionic bath as possible. Nevertheless, we demonstrate that it exhibits non-Markovian effects that have a considerable impact on both the short- and long time dynamics, and thus even for long times it has an effect that cannot be captured by a standard thermodynamic description.
In addition, the simple fermionic gas shares the fact with strongly correlated (critical) systems that its
properties depend only on a few global parameters, in this case the Fermi energy $E_F$ (or the band width $\xi$,
assuming that $\xi \sim E_F$) and the temperature $T$. The obtained results therefore provide also a hint as
to what could be expected as signatures of non-Markovian behavior of correlated systems.
In a similar way a bosonic environment could be investigated.
Goal (\emph{iii}) involves a proof of principle that the non-Markovianity allows for an active manipulation
of the environment. Since the non-Markovianity is driven by quantum fluctuations
we propose to use it as a largely temperature independent way to transport heat out of the environment. Such a method
would thus make it possible to overcome the bottleneck of diverging time scales in demagnetisation cooling
methods which arises because the used relaxation times scale with $1/T$.

The paper is structured as follows. In \cref{sec:Method} we introduce the generalised master equation within the projection operator approach and its formal solution for a particular system. This approach is a controlled expansion in the interaction strength, capturing the full coherent quantum dynamics for non-Markovian regime as well as the long time evolution. Taking into account temporal correlations in the bath we investigate how the dynamics of the spin system is affected if a backaction from the bath is present. In \cref{sec:FermiGas} we derive an analytical solution for the dynamics of the spin system, including the Markovian and non-Markovian decay. In \cref{sec:Cooling} we propose a cooling protocol to utilize the short time dynamics to overcome thermodynamic limitations in the context of cooling an electronic system.

%
\section{Tracking the full time evolution of a quantum system}\label{sec:Method}

Our aim is to determine the time evolution of a spin system coupled to a fermionic environment including the full quantum coherent dynamics. In particular, our work captures the influence of memory effects on the dynamics.
The generic system considered here consists of an impurity spin-1/2, e.g.\ a nuclear spin or a localized paramagnetic electron spin, coupled to an electronic bath.
It is described by the Kondo-type Hamiltonian
\begin{align}
    H= \sum_{k\sigma}\epsilon_kc^\dagger_{k\sigma}  c_{k\sigma} + b_z^\text{el}\sum_j S_{z,j} + b_z^I I_z + A \mathbf{I}\cdot\mathbf{S}_{j=0}.
    \label{eq:H1}
\end{align}
The first term models the electronic environment with annihilation and creation operators $c_{k\sigma}$ and $c^\dagger_{k\sigma}$. The momenta $k$ label all possible states with energy dispersion $\epsilon_k$, and $\sigma$ is the spin index. The second and third term are the Zeeman terms for the electrons and the impurity spin. Here, $\mathbf{I}$ is the impurity spin operator, and $\mathbf{S}_j = \sum_{\sigma,\sigma'} c_{j,\sigma}^\dagger \boldsymbol{\tau}_{\sigma,\sigma'} c_{j,\sigma'}$ are the electron spin operators, written in terms of the real space operators $c_{j,\sigma}$ and the Pauli matrix vector $\boldsymbol{\tau}$. The index $j$ labels the position of the electrons, where for the ease of notation we assume an underlying lattice, but this is not essential for the physics discussed here.
For convenience we will normalise both spin operators to dimensionless $|\mathbf{S}_j|=|\mathbf{I}|=1$.
We will also assume that $\mathbf{I}$ is a spin-$1/2$ operator since this will make the matrix structure of the formalism used below
simpler, but this assumption is of no further importance, and the formalism and the results can easily be extended to larger spins.
Both spin species interact through a contact interaction with strength $A$. A uniform magnetic field $B_z$ is applied along the $z$ axis and we define $b_z^\text{el} = g \mu_B B_z$ and $b_z^I = g_I \mu_I B_z$, with the g-factors $g$ for the electrons and $g_I$ for the impurity spins, $\mu_B$ for the Bohr magneton, and $\mu_I$ for the magnetic moment of the impurity spin.

Although the Hamiltonian in \cref{eq:H1} is of the Kondo-type we emphasize that we do not consider it in the Kondo regime. We focus on small ratios of $A/E_F$, where $E_F$ is the Fermi energy, such that the Kondo temperature, $T_K \sim \exp(-E_F/A)$, is much below any realistic situation. The small ratio $A/E_F$ also allows us to expand the equation of motion to second order with which we thus neglect the higher order spin-flip terms driving the Kondo effect. Furthermore, we explicitly control definite initial values for the impurity spin and focus on the short time dynamics that would precede the development of Kondo correlations in any case. Nonetheless the elementary particle-hole fluctuations underlying the Kondo effect remain active and cause the logarithmic time dependences found below.

In this work we assume that the electron system remains nonmagnetic, $\langle S_z(t)\rangle= 0$. This allows us to simplify the equations in the following while keeping all the relevant physical aspects.
However, we then also explicitly exclude the Knight shift which describes the shift of the resonances of $\mathbf{I}$ by the effective magnetic field created by the paramagnetic polarization $\langle \mathbf{S} \rangle$ \cite{Abragam,Slichter}.
Although the latter is notable for embedded spins in many electron conductors our focus is on the irreversible behavior from quantum and thermal fluctuations, and the Knight shift can effectively be absorbed by shifting the Zeeman field acting on $\mathbf{I}$. Therefore, to maintain clarity in the discussion we prefer to neglect this shift. Nonetheless we shall in the discussion occasionally come back to its influence.

Changing into a rotating frame of reference eliminates the impurity Zeeman term, $H\rightarrow H - b_z^I J_z$, with the total angular momentum $J_z = I_z + \sum_jS_{z,j}$. The spin operators transform as
\begin{align}
    I_\pm&\rightarrow I_\pm e^{\pm i b_z^I t},\quad &&S_{\pm,j}\rightarrow S_{\pm,j} e^{\pm i b_z^I t},\notag\\
    I_z&\rightarrow I_z, \quad&&\ S_{z,j}\rightarrow S_{z,j},
	\label{eq:rot_frame}
\end{align}
where as usual
\begin{equation}
	I_\pm =\frac{1}{2} \left(I_x \pm i I_y\right), \qquad S_{\pm,j} = \frac{1}{2}\left(S_{x,j} \pm i S_{y,j}\right).
\end{equation}
Later we shall also use the notation
\begin{equation}
	I_\uparrow = \frac{1}{2} (\mathbb{1} + I_z), \qquad
	I_\downarrow = \frac{1}{2} (\mathbb{1} - I_z),
\end{equation}
with $\mathbb{1}$ being the identity operator.
The Hamiltonian \cref{eq:H1} in the new reference frame can then be written as
\begin{align}
   H&=\sum_{k,\sigma}\epsilon_k c^\dagger_{k\sigma} c_{k\sigma}+b_z\sum_jS_{z,j}+ A \mathbf{I}\cdot\mathbf{S}_{j=0},
   \label{eq:Hamiltonian}
\end{align}
with $b_z=b_z^\text{el}-b_z^I$.

The full dynamical behavior of the system is encoded in the time evolution of the reduced density matrix of the system
\begin{align}
    \rho_I = \text{Tr}_\text{el}\left[\rho\right],
\end{align}
where $\rho$ is the full density matrix of the system and the environment.
We furthermore define the projection operator $P$ such that $PO = \rho_\text{el}\otimes\text{Tr}_\text{el}[O]$ for any operator $O$, and $Q = \mathbb{1} - P$ as the complement of $P$. The equilibrium density matrix of the electronic system is denoted as $\rho_\text{el}$ in its initial state and  $\text{Tr}_\text{el}$ is the trace over the electronic degrees of freedom. The Nakajima-Zwanzig equation  provides a framework to formulate an equation of motion for the reduced density matrix $\rho_I(t)$ \cite{Breuer,Sauermann}
\begin{align}
    \frac{d}{dt}\rho_I(t)=- i \int_0^t dt'\,\Sigma_I\left(t-t'\right)\rho_I(t').
    \label{eq:master}
\end{align}
In the latter equation the reduced memory kernel
\begin{align}
     \Sigma_I(t-t')=-i\text{Tr}_\text{el}\left[Le^{-iQL(t-t')}QL_\text{int}\rho_\text{el}\right],
    \label{eq:memory}
\end{align}
carries the information of the system's history and its time dependence needs to be carefully taken into account to capture the non-Markovian behavior.
We split the Hamiltonian $H=H_0 + H_\text{int}$ into two parts. The noninteracting part $H_0=~\sum_{k,\sigma}\epsilon_kc^\dagger_{k\sigma}c_{k\sigma} + b_z\sum_jS_{z,j}$ describes the fermionic bath, and $H_\text{int} = A\mathbf{I}\cdot\mathbf{S}_{j=0}$ is the interaction with the spin system, according to \cref{eq:Hamiltonian}.
Based on these Hamiltonians we define in the Nakajima-Zwanzig equation the Liouvillian superoperators $L_0 O = [H_0, O]$ and $L_\text{int} O = [H_\text{int},O]$, where $O$ is any operator, and $L=L_0+L_\text{int}$.

To solve the integro-differential equation of motion, we analyze \cref{eq:master} in Laplace space.
If $f(t)$ is any function of time, the Laplace transform is given by  $\tilde{f}(s)= \int_0^\infty dt\,\exp{(-ts)}f(t)$, for $\text{Re}(s) > 0$,
and its application on \cref{eq:master} leads to
\begin{align}
    \tilde{\rho}_I(s) &=\left(s\mathbb{1}+i \tilde{\Sigma}_I(s)\right)^{-1}\rho_I(t=0),
\end{align}
with the initial condition $\rho_I(t=0)$ and
\begin{align}
    \tilde{\Sigma}_I(s)=-i\text{Tr}_\text{el}\left[L\left(s\mathbb{1}+iQL\right)^{-1}QL_\text{int}\rho_\text{el}\right].
\label{eq:rho_I_Laplace}
\end{align}
Up to this point, the Nakajima-Zwanzig approach is exact for any Hamiltonian that can be split into interacting and noninteracting parts. A detailed derivation can also be found in \cref{app:master}. To make progress we shall assume that $A<E_F$ such that we can expand the memory kernel $\Sigma_I$ in Laplace space.
We stress that this expansion is only perturbative in the system-bath coupling, but nonperturbative in the time domain.
As we will see below we use the expansion of the memory kernel only to facilitate the identification of the decay modes
of the equation of motion in Laplace space, but we then will consider the contribution of the entire ensemble of modes
to the time evolution. With this we go far beyond the Markov approximation or any expansion in the memory time.
Instead the approach shares some aspects of introducing an approximate self-energy term in the Dyson equation for many-body Green's functions, and indeed the memory kernel $\Sigma$ can be seen as equivalent to the self-energy of nonequilibrium many-body theory.
Therefore, it preserves the possibility of the superposition of the infinite number of quantum fluctuations which govern the non-Markovian behavior. The latter is encoded in the pole structure of the correlators in Laplace space and analyzing those presents a systematic way of fully including the non-Markovian contribution which will be discussed in detail later on.
It is worth noting that the formalism is independent of the nature of the interaction or the actual structure of the bath and can be applied equally to fermionic or bosonic environments.

The approach provides a controlled expansion of the memory kernel in the interaction while keeping all the information about the system's past.
Within the Born approximation the correction of the memory kernel is quadratic in the interaction term $L_\text{int}$, and is thus quadratic in the coupling strength $A$
\begin{align}
    \tilde{\Sigma}^\text{Born}_I(s) = -i\text{Tr}_\text{el}\left[L_\text{int}\left(s\mathbb{1} + QL_0\right)^{-1}QL_\text{int}\rho_\text{el}\right].
\end{align}
Choosing the impurity spin as a basis, i.e.\ the space spanned by the spin operators $\{I_\uparrow,I_\downarrow,I_-,I_+\}$, allows us to decompose any operator into a part that only acts on the bath and a part that only acts on the system. Within this basis the reduced density matrix decomposes as
\begin{equation}
	\rho_I = \rho_\uparrow I_\uparrow + \rho_\downarrow I_\downarrow + \rho_- I_- + \rho_+ I_+,
\end{equation}
such that $\langle I_\beta \rangle = \text{Tr}_I\left[ I_\beta \rho_I\right] = I \rho_\beta$, where $\beta = \uparrow, \downarrow, -, +$ and $\text{Tr}_I$ is the trace over the system's degrees of freedom.

The superoperators can then be represented by $4~\times~4$ matrices acting on the reduced density matrix vector $(\rho_\uparrow,\rho_\downarrow,\rho_-,\rho_+)^T$ \cite{LossCoish}.
In the following, we use square brackets $[O]$ to denote this matrix representation of a superoperator $O$.
The memory kernel $\left[\tilde{\Sigma}_I^\text{Born}(s)\right]$ in its matrix representation takes the form
\begin{align}
    \left[\tilde{\Sigma}^\text{Born}_I(s)\right]=
    \begin{pmatrix}
            F_1&-F_2&0&0\\
            -F_1&F_2&0&0\\
            0&0&F_-+F_z&0\\
            0&0&0&F_++F_z
    \end{pmatrix}.
    \label{eq:main_memory}
\end{align}
The entries of the memory kernel are the Laplace transforms of spin-spin correlation functions $F_j(s)$. The derivation of \cref{eq:main_memory} including the spin-spin correlators $F_j$ can be found in \cref{app:memory}.

Since the Hamiltonian $H_0$ describing the electronic environment is spin conserving the matrix entries in \cref{eq:main_memory} that do no conserve the electron spin are zero, and the remaining terms $F_{1,2,\pm, z}$ describe variants of $\langle S_\mp S_\pm \rangle$
or $\langle S_z S_z \rangle$ electron spin correlators (see \cref{app:memory}). The memory kernel $\tilde{\Sigma}_I$ takes into account electronic quantum fluctuations induced by an excitation of the impurity spin. Because of the Pauli principle such fluctuations are dominated by particle-hole excitations at the Fermi level. These propagate through the bath and act back onto the impurity spin, creating an effective time-retarded coupling of the impurity spin with itself. Thus, before reaching the thermodynamic equilibrium regime the impurity spin is correlated with its initial state.

With the memory kernel in the matrix representation of \cref{eq:main_memory} the solution of \cref{eq:rho_I_Laplace} is a simple matrix inversion. The application of the inverse Laplace transform on this solution then provides the full time evolution of the density matrix $\tilde{\rho}_I(s) \rightarrow \rho_I(t)$. For the longitudinal and transverse components, $\rho_z(t) = \rho_\uparrow(t) - \rho_\downarrow(t)$ and $\rho_\pm(t)$, this yields
\begin{align}
    \rho_z(t)&=\int_{-i\infty +\lambda}^{i\infty + \lambda} \frac{ds}{2\pi i}\,e^{st}\,\frac{s\rho_z(t=0)+iF_2(s)-iF_1(s)}{s\left[s+iF_1(s)+iF_2(s)\right]},
    \label{eq:rhoz}\\
    \rho_\pm(t)&=\int_{-i\infty +\lambda}^{i\infty+\lambda}\frac{ds}{2\pi i}\,e^{st}\,\frac{\rho_\pm(t=0)}{s+iF_\pm(s)+iF_z(s)},
    \label{eq:rhopm}
\end{align}
where $\lambda$ is a real number such that all singularities of the integrand lie to the left of the integration contour. Notice that at this order of approximation the equations for $\rho_z$ and $\rho_\pm$ decouple. Any cross dependence would require a further expansion of the memory kernel which is beyond our current interest.
%
\begin{figure}[t]
    \centering
    \includegraphics{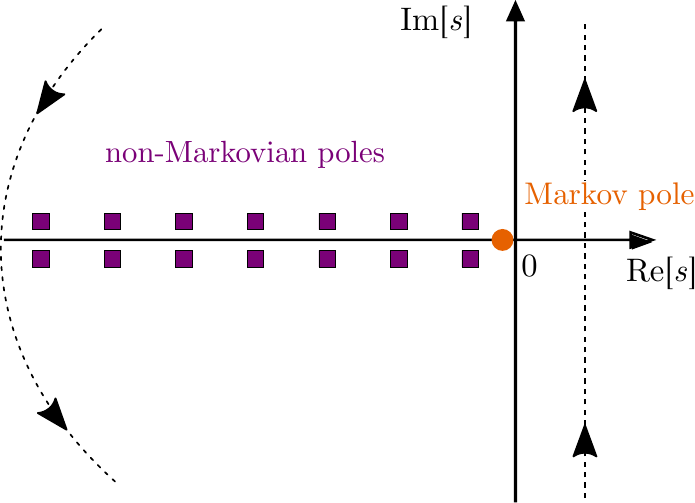}
    \caption{Sketch of the pole structure in Laplace space. The orange point marks the Markov pole with finite negative real part giving the exponential decay. The non-Markovian poles in purple contribute to the dynamics at all temperatures. In the zero field limit $b_z^I=0$, the poles shift onto the real axis. The black dashed lines indicate the semi-circular Bromwich contour for the inverse Laplace transform.}
    \label{fig:contour}
\end{figure}
%
The quantities $\rho_{z,\pm}(t=0)$ are the initial states described by the corresponding density matrix.
The full time evolution of the reduced density matrix is determined by the complex contour integration along the Bromwich contour shown in \cref{fig:contour}. The dynamical behavior is dominated by the location of the integrand's poles or branch cuts in Laplace space. Such singular structures in Laplace space correspond directly to a collective response of the physical system, and in much of our following discussion we shall read off directly the physical consequences from the position of the poles.

In general we must distinguish between two types of singular structures. The Markov approximation consists of neglecting any $s$ dependence of the $F_{1,2,\pm,z}(s)$ functions and picking up the simple pole at $s = -i [F_1(0)+F_2(0)]$ in \cref{eq:rhoz} or $s= -i [F_\pm(0)+F_z(0)]$ in \cref{eq:rhopm}. If we keep the full $s$ dependence there remains a pole near these values which we shall continue calling the \emph{Markov pole}. From the residuum theorem this pole leads to a purely exponential decay of the initial state, characteristic for the memoryless Markov behavior. The time scale for this decay is set by the residue of the Markovian pole which corresponds to the spin-lattice relaxation time $T_1$ for the longitudinal decay and the spin-spin relaxation or decoherence time $T_2$ for the transverse decay.

In addition to the Markov pole, the full $s$ dependence of the denominators can lead to further singularities. At zero temperature, these can take the form of branch cuts but at any finite temperature these cuts split into infinite sequences of poles with a spacing proportional to the temperature. We shall call these poles the \emph{non-Markovian poles}. Although each pole leads to an exponential decay, the superposition of an infinity of them results in an overall decay that is slower than exponential and this represent the non-Markovian memory effect from the backaction of the bath. This memory effect is eventually suppressed by thermal fluctuations so that the non-Markovian decay is only substantial on times shorter than the memory time $\tau_T \sim \hbar/k_B T$ set by the temperature. The crossover behavior with respect to the time $\tau_T$ can indeed be understood as follows: The separation between the non-Markovian poles is proportional to the temperature $T$. At short times, a large number of these poles contribute to the integral and the time evolution is nonexponential. In particular, in the limit $T\to0$ the poles merge to a branch cut, causing the typical algebraic decay of a fermionic response. At larger times, at $T>0$, the number of poles contributing with a non-negligible amplitude shrinks until at $t>\tau_T$ the decay is essentially described by the exponential arising from the pole closest to the Markov pole. Since the non-Markovian decay is then faster than the Markovian decay the non-Markovian behavior becomes invisible when $t$ passes the time set by the thermal fluctuations.

This description of how the singular structure of the denominator of Eqs. \eqref{eq:rhoz} and \eqref{eq:rhopm} shapes the time range over which non-Markovian effects are notable is general, and the same qualitative behavior will occur for any bath. To become quantitative we need to focus on a specific type of bath.

\section{Memory effects in a simple metal}\label{sec:FermiGas}
The existence of the rich pole structure discussed above indicates already that the non-Markovian behavior is a universal feature and does not depend on specific system and bath types. To illustrate this generality we show that the effect already appears in the simplest possible fermionic bath, a Fermi gas. Although a simple system the latter fully encodes a many-body response since the spin fluctuations are strongly constrained by the Pauli principle. As such a free Fermi gas is already prototypical to more involved correlated electron systems and has the advantage that all calculations can be performed explicitly.
The full system is described by the Hamiltonian of \cref{eq:Hamiltonian}. For simplicity we assume that the system is not magnetized, $\langle S_{x,y,z} \rangle = 0$ even if a small magnetic field is applied. This is a reasonable assumption for metals in which $E_F$ largely exceeds the Zeeman energy but this also excludes explicitly the Knight shift. As mentioned earlier this makes the discussion of the irreversible behavior of the spin dynamics more transparent, and a later inclusion of the Knight shift is straightforward.

To start, we shall assume that the magnetic field is zero. Due to the full $SU(2)$ symmetry of the combined system, we then have
\begin{align}
    F_1(s) &= F_2(s)=F_\pm(s) = F_z(s),
\end{align}
which allows us to express all correlators in terms of $F_z$ (see \cref{app:memory}),
\begin{align}
\label{eq:F_z}
    F_z(s) =\frac{A^2}{2i} \int_0^\infty dt \,e^{-t s} \langle\left\{ S_{z,j=0}(t), S_{z,j=0}(0)\right\}\rangle.
\end{align}
Here $\langle O\rangle$ denotes the average over the electronic equilibrium state, $\text{Tr}_\text{el} [O\rho_\text{el}]$, for any operator $O$ and $\{\cdot,\cdot\}$ is the anti-commutator. A detailed derivation of the analytic expression for the spin-spin correlators is presented in \cref{app:correlator}.
Since the electronic Hamiltonian is quadratic, the spin-spin correlators decouple into simple fermionic expectation values and \cref{eq:F_z} reduces essentially to integrations over Fermi functions and exponentials. For a conventional metal with an approximately constant density of states about the Fermi energy, the only relevant parameter for the universal low-energy behavior is the electron temperature $T$. This results in two distinct regimes for the spin-spin correlator $F_z(s)$. In the region where $\text{Re}(s)<k_BT/\hbar$, thermal fluctuations determine the system's dynamics. On the other hand, for $\text{Re}(s)>k_BT/\hbar$, quantum fluctuations are the dominant contributions and memory effects arise from particle-hole fluctuations about the Fermi level.
These fluctuations are universal in that they depend only on the existence of the Fermi surface, and we evaluate the correlators of
\cref{eq:F_z} under the aspect of this universality. However, the general treatment is not limited to this type of environment and as long
as a computation of the correlators $F_j(s)$ is possible the method can be applied to any other bath as well.

The correlators $F_j(s)$ are proportional to the interaction $A^2$ due to the Born approximation of the memory kernel $\tilde{\Sigma}_I(s)$. In the following we use a small expansion parameter $\alpha$ set by the interaction strength $A$ and the density of states at the Fermi level $\nu_0$
\begin{align}
    \alpha = \left(\nu_0A\right)^2.
\label{eq:alpha}
\end{align}
Generally, $\nu_0 \sim E_F^{-1}$ with $E_F$ being the Fermi energy, such that $\alpha \sim \left( A /E_F\right)^2$. We assume that $\alpha$ is small; in fact, we already assumed the interaction strength $A$ between system and environment is small, such that the Born approximation for the equation of motion is justified. Compared to the Kondo problem we are thus in a regime, where the Born approximation does not break down and the effective coupling does not diverge.
The spin-spin correlation functions for a bath modelled as a Fermi gas up to $O(\alpha)$ (see \cref{app:correlator} for details) can be written as
\begin{align}
    F_z(s) = i \alpha \bigl[ F(s) + G(s) \bigr],
\end{align}
with
\begin{align}
F(s) &= -\pi k_B T  \notag\\
        &+s\left[\ln{\left(\frac{2\pi k_B T}{i\xi}\right)}+\psi\left(1+\frac{s}{2\pi k_B T}\right)\right],
		\label{eq:mainf}
\\
G(s) &= -\pi k_B T  \notag\\
        &+s\left[\ln{\left(\frac{2\pi k_B T}{-i\xi}\right)}+\psi\left(1+\frac{s}{2\pi k_B T}\right)\right],
		\label{eq:maing}
\end{align}
where $k_B$ is the Boltzmann constant, $\psi(z)=~[\frac{d}{dz}\Gamma(z)]/\Gamma(z)$ is the digamma function with $\Gamma(z)$ being Euler's Gamma function, and $\xi \sim E_F$ is a high energy cutoff on the order of the Fermi energy (or the bandwidth). The details on the derivation can be found in \cref{app:correlator}. Although $F(s)$ and $G(s)$ are almost identical, we shall need both of them to unambiguously express the correlators at nonzero magnetic field (see below).

Equations \eqref{eq:mainf} and \eqref{eq:maing} are universal in that they capture exactly the low-energy fluctuations of the electronic excitations. All high-energy fluctuations depending on the nonuniversal details of the band structure are absorbed in the cutoff $\xi$ within the logarithms. The latter contribute only weakly to the dynamics of the spin $I$ at very short times $t < \hbar/\xi$. This allows us to focus on the universal behavior at longer times in the following, fully encoded in $F(s)$ and $G(s)$.

The existence of quantum critical correlations in $F_z(s)$ is best visible in the limit $T \to 0$. Then the digamma function $\psi$ has the asymptotes of a logarithm and the $T$ dependence in the logarithm cancels, which leads to
\begin{align}
    F_z(s) \sim 2i\alpha  s \ln{\left(\frac{s}{\xi}\right)}.
    \label{eq:FzT0}
\end{align}
This result also conveniently bypasses a direct $T=0$ calculation, which would be much more involved. A logarithmic behavior of spin-spin correlators as in \cref{eq:FzT0} signals also the existence of an underlying Kondo effect. However, as mentioned earlier, by maintaining a small $\alpha$ and preparing a well-defined state $\rho_I(t=0)$, the onset of Kondo physics even at very low temperatures is either absent or would occur at rather long times. Therefore, even in then regime where the correlation functions reflect the effective $T \to 0$ limit through their logarithmic form, the temperatures are still high enough to neglect any Kondo physics.

The coherence disappears with increasing temperature, and indeed at large $k_B T \gg \text{Re}(s)$ the digamma function $\psi$ tends to a constant that can be absorbed in the cutoff $\xi \to \xi'$, and we find that
\begin{align} \label{eq:F_z_b0}
    F_z(s)\sim - 2\alpha i \pi k_B T + 2i\alpha s\ln{\left(\frac{2\pi k_B T}{\xi'}\right)}.
\end{align}
The $s$ independent term corresponds to the Markov approximation and will thus cause the standard exponential decay. The second term provides the further renormalization from the electron fluctuations. In this high-temperature limit, this correction is proportional to $s$ and thus causes only a reduction of the Markovian decay amplitude. But for $\text{Re}(s) > k_B T$, which means for times $t < \hbar / k_B T$, the effects caused by the quantum correlations cause a significant deviation from the Markovian exponential behavior.

Before evaluating this dynamics through \cref{eq:rhoz,eq:rhopm}, we should recall that this $F_z(s)$ incorporates the dynamics at zero field. An investigation for $b_z^I \neq 0$ requires us to reintroduce field-dependent phase factors in the correlation functions, leading to shifts of the arguments of $F(s)$ and $G(s)$.
From \cref{eq:app1,eq:app2,eq:app3,eq:app4,eq:app5}, we have
\begin{align}
    F_1(s) &= i\alpha \left[F\left(s-ib_z^I\right) + G\left(s+ib_z^I\right)\right],\label{eq:mainf1}\\
    F_2(s) &= i\alpha \left[F\left(s+ib_z^I\right) + G\left(s-ib_z^I\right)\right],\label{eq:mainf2}\\
    F_-(s) &= F_z\left(s+ib_z^I\right),\label{eq:mainfm}\\
    F_+(s) &= F_z\left(s-ib_z^I\right),\label{eq:mainfp}\\
    F_z(s) &= F_z(s)\label{eq:mainfz},
\end{align}
with $F_z(s)$ as given in \cref{eq:F_z_b0}.
The full time evolution of the reduced density matrix is obtained through \cref{eq:rhoz} and \cref{eq:rhopm}. In the latter equations, the contour integration is evaluated through the zeros of the denominators $s + i F_1(s)+iF_2(s)$ and $s + i F_\pm(s) + i F_z(s)$, respectively. Since $F_{z,1,2,\pm}$ are proportional to $\alpha$, there are two types of zeros. First, there is a zero at $s \sim \alpha$, leading to what we have called the \emph{Markov pole} in the previous section. Second, there is a sequence of zeros near the poles of the digamma function $\psi$. Indeed, in the vicinity of a pole, $\psi$ can raise to $\psi \sim 1/\alpha$ such that $\alpha \psi \sim 1$, which in turn can compensate for the remaining terms in the denominator. Since the digamma function $\psi(z)$ has poles at $z=0,-1,-2,\dots$, this leads to a dynamics governed by the quantum fluctuations of the Fermi gas expressed by the $\psi$ term, and above we called these poles the \emph{non-Markovian poles}. The resulting behavior will be calculated explicitly in the next subsections.

For $\rho_z$, it should also be noted that, in \cref{eq:rhoz}, there is a further pole at $s=0$. This corresponds to the equilibrium value $\langle I_z \rangle = I \rho_z(t \to \infty)$ and we will include it in the discussion of the Markov behavior.


\subsection{Markovian decay}

\begin{figure}
    \includegraphics{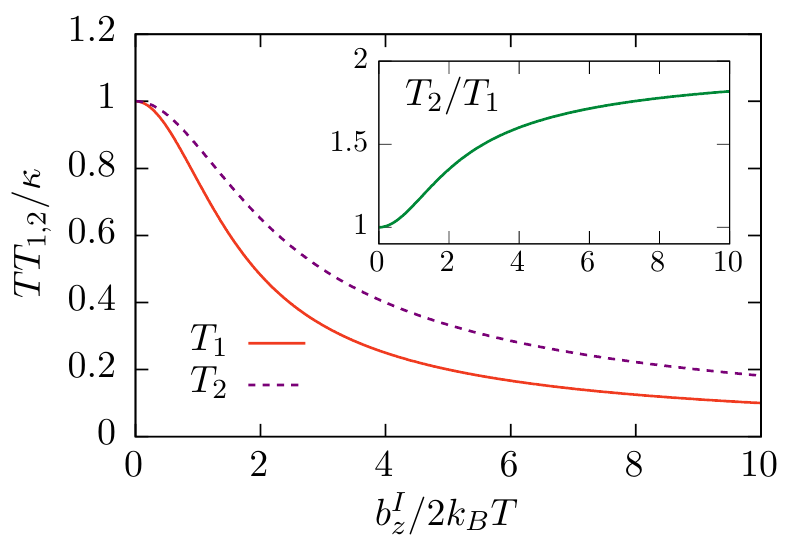}
    \caption{
    Scaling behavior of the $T_1$ and $T_2$ times given in \cref{eq:Korringa,eq:T2} as a function of the ratio of magnetic field to temperature, $b_z^I/2k_B T$. Plotted is $T T_{1,2} / \kappa$, where $T$ is temperature and $\kappa = \hbar/4\alpha \pi k_B$ is the Korringa constant. At $b_z^I=0$ the $SU(2)$ symmetry imposes $T_1=T_2$ and we have $T_{1,2} = \kappa T$. At nonzero field, the equality of $T_1$ and $T_2$ is lifted and both times become shorter. In the limit of very large $b_z^I$, we obtain $T_2 = 2 T_1$. The inset shows how the ratio $T_2/T_1$ evolves for the same $b_z^I/2k_BT$ values as in the main plot.
    \label{fig:T1T2}}
\end{figure}

The Markov approximation neglects the $s$ dependence of the spin-spin correlation functions \cref{eq:mainf1,eq:mainf2,eq:mainfm,eq:mainfp,eq:mainfz} by assuming that the system does not have a memory time, i.e.,  the equation of motion \cref{eq:master} is time local. The corresponding Markov pole is on the order of $s \sim \alpha$ and has a negative real part. This leads to an exponential decay of the density matrix components governed by the decay times $T_{1,2} = -1/\mathrm{Re}(s)$, with the relaxation time $T_1$ and the dephasing time $T_2$ describing the evolution of $\rho_z$ and $\rho_\pm$ respectively.

We have thus to solve for $s + iF_1(s)+ iF_2(s)= 0$ and $s + iF_+(s)+ iF_-(s)= 0$ for general $b_z^I$ under the condition $s \sim \alpha$. As a general strategy, we shall profit from the smallness of $\alpha$ to expand the position $s$ of the poles and their residues to order $\alpha$. However, to capture the full time dynamics nonperturbatively, we must not expand the time-dependent exponentials $e^{s t}$. The details of this calculation for the Markov part are given in \cref{app:Markov}.
From the real part of the corresponding Markov poles, we obtain the decay times
\begin{align}
    T_1&=\frac{\hbar}{2\alpha\pi b_z^I}\tanh\left(\frac{b_z^I}{2 k_B T}\right),
    \label{eq:Korringa}\\
    T_2&=\frac{\hbar\tanh\left(\frac{b_z^I}{2 k_B T}\right)}{2\alpha \pi k_B T\tanh\left(\frac{b_z^I}{2 k_B T}\right)+\alpha\pi b_z^I}\label{eq:T2}.
\end{align}
In the zero field limit $b_z^I=0$, these expressions become
\begin{align}
    T_1 = T_2 = \frac{\hbar}{4\alpha\pi k_B T}, \label{eq:T1_T2_b0}
\end{align}
which recovers the Korringa relation $T T_1 = \kappa$ of Fermi liquids \cite{Korringa,Moryia,Abragam,Slichter}, with $\kappa = \hbar/4\alpha \pi k_B$ being the Korringa constant. Expressed in $\kappa$, \cref{eq:Korringa} coincides with the magnetic field dependent decay times found in the literature \cite{Pobell,Slichter,Abragam}.
Furthermore, the equality $T_1=T_2$ is the consequence of the $SU(2)$ symmetry of the system. This is in contrast to systems with a broken $SU(2)$ symmetry, in which $T_2=2T_1$ can be achieved \cite{Golovach,Borhani,Slichter,Abragam}. The latter result can also obtained in the present case by introducing symmetry breaking through the magnetic field. Indeed, in the limit $b_z^I \gg k_B T$, \cref{eq:Korringa,eq:T2} become $T_1=\hbar/2\alpha\pi b_z^I$ and $T_2=\hbar/\alpha\pi b_z^I$, and we recover $T_2 = 2T_1$.
In \cref{fig:T1T2}, we show the universal behavior of $T T_{1,2}/\kappa$ as a function of $b_z^I/2 k_B T$. It is notable that $T_2$ fulfils for any nonzero field $T_1 < T_2 < 2T_1$ and that the upper limit $T_2=2T_1$ is reached only rather slowly for large $b_z^I$ values, which is best seen through the ratio $T_2/T_1$ shown in the inset of \cref{fig:T1T2}.

From \cref{app:Markov}, we obtain then the Markovian part of the time evolution
\begin{align}
    &\rho^\text{M}_z(t)=\rho_z^\text{eq}\notag\\&+\frac{\left[\rho_z(t=0)-\rho_z^\text{eq}\right]e^{-t/T_1}}{1-4\alpha\left\{\ln{\left(\frac{2\pi k_B T}{\xi}\right)}+\text{Re}\left[\psi\left(1+\frac{ib_z^I}{2\pi k_B T}\right)\right]\right\}},\label{eq:markovz}\\
    &\rho^\text{M}_\pm(t)=\frac{\rho_\pm(t=0)e^{-t/T_2}e^{it\omega_\pm}}{1-2\alpha\left[2\ln{\left(\frac{2\pi k_B T}{\xi'}\right)}+\psi\left(1\mp\frac{ib_z^I}{2\pi k_B T}\right)\right]}\label{eq:markovpm},
\end{align}
where $\rho_z^\text{eq}$ is the paramagnetic equilibrium magnetization corresponding to complete thermalization with the electronic bath,
\begin{equation}
	\rho_z^\text{eq} = - \tanh\left(\frac{b_z^I}{2\pi k_B T}\right),
\end{equation}
and where the transverse component contains a spin precession described by the frequency
\begin{align}
\omega_\pm = \mp \frac{2\alpha b_z^I}{\hbar}\left\{\ln\left(\frac{2\pi k_B T}{\xi'}\right)+\text{Re}\left[\psi\left(1\mp\frac{i b_z^I}{2\pi k_B T}\right)\right]\right\}.
\end{align}
In the latter expressions, $\xi'$ is an inessential renormalization of $\xi$ obtained by absorbing a constant $\psi(1)/2$ in the cutoff. Through a similar further shift of the cutoff to $\xi''$, we can rewrite the zero field expressions as
\begin{align}
	\rho^\text{M}_{z,\pm}(t)= \frac{\rho_{z,\pm}(t=0) e^{-t/T_{1,2}}}{1-4\alpha \ln\left(\frac{2\pi k_B T}{\xi''}\right)}, \label{eq:markovb0}
\end{align}
with $\xi'' = \xi \exp(-\psi(1))$.

Notice that in contrast to the standard Markovian decay we have kept an $O(\alpha)$ correction in the amplitudes of \cref{eq:markovz,eq:markovpm,eq:markovb0}. This results from keeping the $s$ dependence in $F_{z,1,2,\pm}(s)$ instead of setting in the latter functions $s=0$ and solving, for instance, $s+iF_1(0)+iF_2(0)=~0$. As a consequence, the weight of the amplitudes is reduced from 1 to $[1- O(\alpha)]$. This effect alone already indicates the presence of the further non-Markovian decay terms and that the latter have an amplitude of order $\alpha$.

\begin{figure}
    \includegraphics{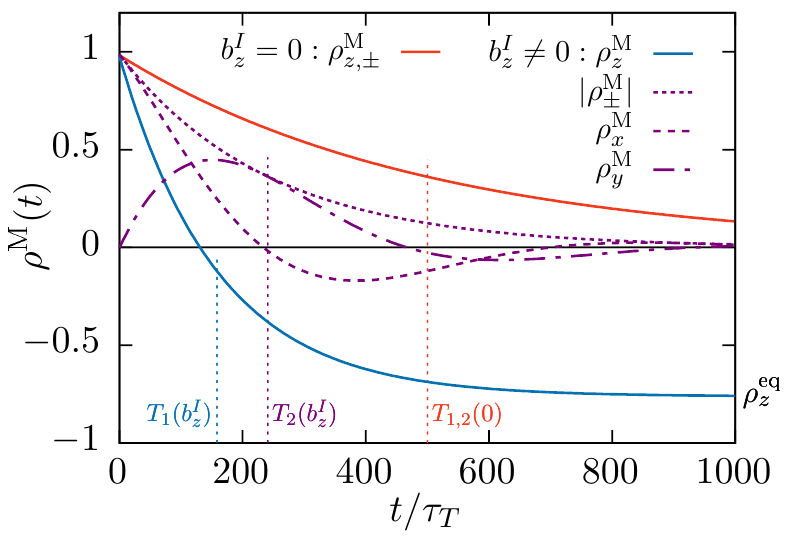}
    \caption{
	Decay of the reduced density matrix as a function of time. The solid orange curve shows the decay at zero field $b_z^I=0$ at which $\rho_{z}^\text{M}=\rho_\pm^\text{M}$ and $T_1=T_2$. The chosen parameters are $\alpha=10^{-3}$ and $k_B T=\xi/200$, and time is plotted in units of $\tau_T=\hbar/2\pi k_B T$. The further curves show the time dependence at rather large field $b_z^I = 4 \pi k_B T=2\hbar/\tau_T$ at which the $T_{1,2}$ times become shorter and $\rho_\pm^\text{M}$ becomes oscillating through $\exp(i \omega_\pm t)$. Shown are $\rho_z^\text{M}$ (solid, blue) as well as $|\rho_\pm^\text{M}|, \rho_x^\text{M} = \text{Re}[\rho_\pm]$, and $\rho_y^\text{M} = \pm \text{Im}[\rho_\pm^\text{M}]$ (various dashes, purple). Notice that for $b_z^I>0$ the spin $\rho_z^M(t=0)=+1$ is initialized against the magnetic field and the equilibrium value is therefore negative, $\rho_z^\text{eq}<0$.
    \label{fig:Markov}}
\end{figure}

The Markovian decay according to \cref{eq:markovz,eq:markovpm} is shown in \cref{fig:Markov} for initial conditions $\rho_z(t=\hbar\xi^{-1})= 1$ and $\rho_\pm(t=\hbar\xi^{-1}) =1$, respectively. In the zero-field limit, the evolution of $\rho_z(t)$ and $\rho_\pm(t)$ are identical (orange curve). For $b_z^I>0$, both $T_1$ and $T_2$ times decrease. In the figure, we show the decay for a rather large field $b_z^I = 2 \pi k_B T$ in which $\rho_z^\text{eq}=-\tanh(1) \approx -0.76$ is significant (blue curve), and the effect of $\exp(i \omega_\pm t)$ in $\rho_\pm^\text{M}$ becomes visible since $\omega_\pm \sim 1/T_2$ (purple dashed curves).
Such oscillations renormalize the normal precession from the spin in the magnetic field that we have removed by going to the rotating frame.
As such, their role is similar to the Knight shift, which we have neglected in the present treatment but which also adds to the eigenfrequency. But in contrast to the Knight shift, the $\omega_\pm$ have a strong nonlinear magnetic field dependence. They are only proportional to $b_z^I$ at low fields but for $b_z^I > 2 \pi k_B T$ are strongly sub-linear and eventually, at very strong fields, change sign.

%
\subsection{Non-Markovian contributions}
Apart from the isolated Markov pole, we find an infinite sequence of poles close to the singularities of the spin-spin correlators \cref{eq:mainf1,eq:mainf2,eq:mainfm,eq:mainfp,eq:mainfz}. While for the Markov poles $s \sim \alpha$, the $s$ values of these new poles are to leading order independent of $\alpha$. Expressions like $s + i F_1(s) + i F_2(s)$ can then only be zero if a divergence of order $1/\alpha$ compensates their small $\alpha$ amplitude. This means that the $s$ have to lie near the singularities of $F$ and $G$. For $F(s)$ and $G(s)$ as given by \cref{eq:mainf,eq:maing},  the divergences are governed by the singularities of the digamma functions $\psi\left(1+s/2\pi k_B T\right)$ and $\psi\left(1+(s\pm i b_z^I)/2\pi k_B T\right)$. The digamma function $\psi(1+z)$ has simple poles at $z=-1,-2,\dots$ and in their vicinity we find the necessary $\psi(1+z) \sim 1/\alpha$ behavior. This allows us to systematically expand the denominators of \cref{eq:rhoz,eq:rhopm} and pick up the residues to order $\alpha$. The detailed calculation is done in \cref{app:nonmarkovz}. Remarkably, we can do the summation over the residues exactly and the resulting non-Markovian contributions to $\rho_z$ and $\rho_\pm$ are given by
\begin{align}
	\rho_z^{\text{nM}}(t)
	&= -4\alpha e^{-4\alpha t/\tau_T}\ln{\bigl(1-e^{-t/\tau_T}\bigr)} \notag \\
	& \ \ \ \times h(t,b_z^I) [\rho_z(t=0)-\rho_z^\text{eq}],
	\label{eq:nM_z}
\\
	\rho_\pm^{\text{nM}}(t)
	&= -4\alpha e^{\pm i b_z t/2\hbar} e^{-4\alpha t/\tau_T}\ln{\bigl(1-e^{-t/\tau_T}\bigr)} \notag \\
	&\ \ \ \times h(t,b_z^I/2) \rho_\pm(t=0),
	\label{eq:nM_pm}
\end{align}%
%
%
\begin{figure}[t!]
    \includegraphics{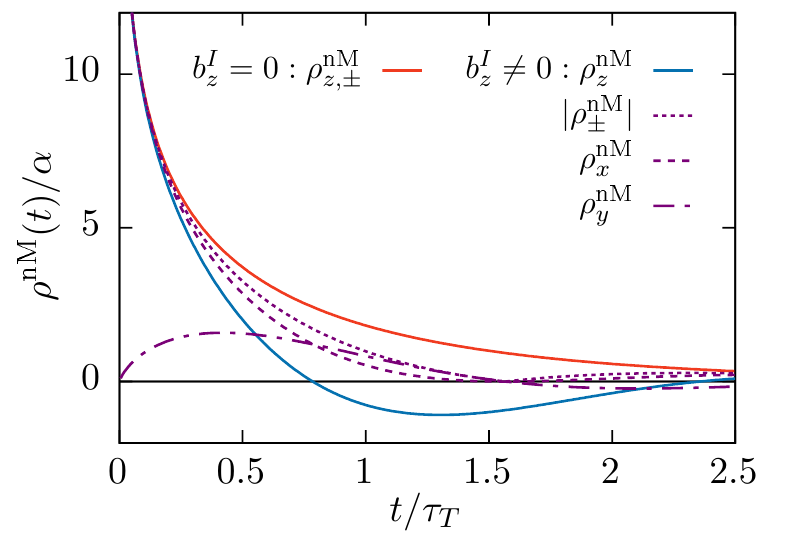}
    \caption{Plot of the non-Markovian evolution of $\rho_z$ and $\rho_\pm$ according to \cref{eq:nM_z,eq:nM_pm}. The solid orange curve shows $\rho_z^\text{nM}=\rho_\pm^\text{nM}$ at zero magnetic field. These curves remain almost unchanged for fields $b_z^I$ up to a significant fraction of $k_B T$, although for $b_z^I > 4 \alpha \pi k_B T$ slow oscillations appear due to the function $h$. But the oscillations are strongly damped and only visible for strong magnetic fields. For illustration, we choose a strong field $b_z^I = 4 \pi k_B T = 2 \hbar/\tau_T$. The solid blue curve then shows $\rho_z^\text{nM}$ and the various purple dashed lines $|\rho_\pm^\text{nM}|, \rho_x^\text{nM} = \text{Re}[\rho_\pm^\text{nM}]$, and $\rho_y^\text{nM} = \pm \text{Im}[\rho_\pm^\text{nM}]$. As in \cref{fig:Markov}, we have chosen $\alpha = 10^{-3}$ and $k_B T = \xi/200$.
    \label{fig:NMrhop}}
\end{figure}%
%
with $\tau_T = \hbar/2\pi k_B T$ being the thermal time and $h(t,b_z^I)$ being the function
\begin{align}
	&h(t,b_z^I)
	=
	\cosh{\left(\frac{t}{\hbar}\sqrt{(4\alpha\pi k_B T)^2-(b_z^I)^2}\right)}\notag\\
    &-\frac{4\alpha\pi k_B T}{\sqrt{(4\alpha \pi k_B T)^2-(b_z^I)^2}}\sinh{\left(\frac{t}{\hbar}\sqrt{(4\alpha\pi k_B T)^2-(b_z^I)^2}\right)}.
	\label{eq:h}
\end{align}%
%
\begin{figure*}[t!]
    \includegraphics{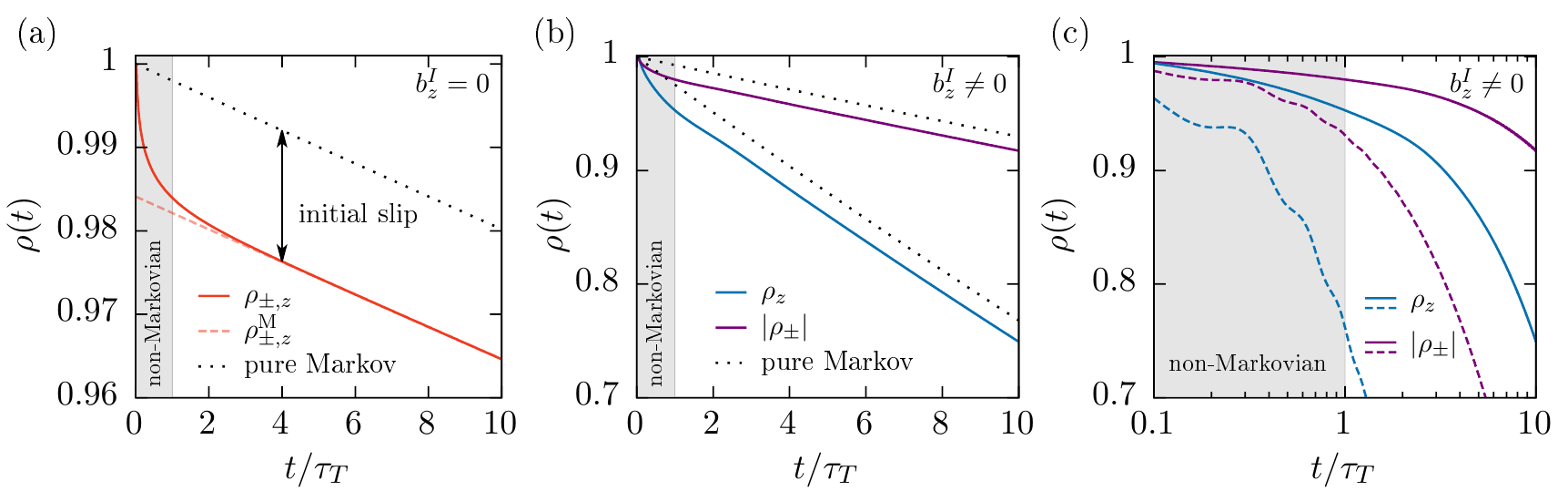}
    \caption{Full time evolution of the reduced density matrix components $\rho_z$ and $\rho_{\pm}$.
	(a) Evolution of $\rho_{z}(t)=\rho_{\pm}(t)$ (solid orange curve) at zero magnetic field, $b_z^I=0$, in comparison with the standard pure Markov approximation $\rho_{z,\pm}(t) = e^{-t/T_{1,2}}$ (dotted black curve). The non-Markovian behavior is dominant in the grey area at times $t<\tau_T=\hbar/2\pi k_B T$, after which $\rho_{z,\pm}(t)$ cross over in an exponential decay parallel to the pure Markovian exponential. The offset is marked as an `initial slip' and is measurable through the extrapolation of the Markovian behavior to time $t=0$ (dashed pale-orange line).
	(b) Evolution of $\rho_z$ (blue) and $|\rho_\pm|$ (purple) at nonzero field $b_z^I = 4\pi k_B T = 2 \hbar/\tau_T$, in comparison again with the pure Markovian exponential decays $e^{-t/T_{1,2}}$ (dotted black lines).
	(c) Evolution on a logarithmic time scale. The solid curves are identical to those in panel (b). For the dashed curves, a very large field of $b_z^I = 40 \pi k_B T = 20 \hbar/\tau_T$ was chosen to illustrate the oscillations induced through the function $h(t,b_z^I)$ in both $\rho_z$ and $|\rho_\pm|$.
	In all plots, the parameters are $\alpha=10^{-3}$ and $k_B T=\xi/200$ as in \cref{fig:Markov,fig:NMrhop} and the spurious divergences of the non-Markovian behavior as $t \to 0$ have been cut off by starting the curves at a time $t \sim \hbar/\xi$.
    \label{fig:fulldecay}}
\end{figure*}%
%
Notice that since $\ln(1-e^{-t/\tau_T}) \sim e^{-t/\tau_T}$ for $t > \tau_T$ the characteristic decay time for these non-Markovian parts is $\tau_T$ and not $T_{1,2}$ which would be by $1/\alpha$ longer. This is not surprising as temperature fluctuations erase any memory effect. However, it is notable is that we have obtained an explicit prescription of the non-Markovian quantum behavior within the memory time. At low magnetic fields, the scale-free nature of the Fermi gas becomes manifest in that temperature is the only scaling parameter for the dynamic behavior. Larger magnetic fields introduce a further scale to the system, but we see that their effect is limited, entering in a nontrivial way only through $h(t,b_z^I)$. The latter causes either only a small renormalization of the overall decay for $b_z^I < 4 \alpha \pi k_B T$, or the addition of further, rather slow oscillations for $b_z^I > 4 \alpha \pi k_B T$. These oscillations are similar to the Knight-shift-type effect of $\omega_\pm$ for the Markov contribution but they appear here for both $\rho_\pm^\text{nM}$ and $\rho_z^\text{nM}$ (see \cref{fig:NMrhop}), and with different onset fields and frequencies due to the dependence through $h(t,b_z^I/2)$ and $h(t,b_z^I)$, respectively. At zero field, the evolution of $\rho_z^\text{nM}$ and $\rho_\pm^\text{nM}$ coincide (\cref{fig:NMrhop}, orange curve). It should finally be noted that \cref{eq:h} has been derived under the assumption that the maximally considered $b_z^I$ does not excessively exceed $2\pi k_B T$. Nevertheless, we expect that \eqref{eq:h} remains a good approximation even in the limit $b_z^I \gg 2\pi k_B T$ because corrections to the derivation done in \cref{app:nonmarkovz} can only weakly renormalise the position of the non-Markovian poles.

%
%
%
\subsection{Total decay and limits}

The full time evolution of the reduced density matrix is the sum of the Markovian and the non-Markovian contributions, given for $\rho_z$ by \cref{eq:markovz,eq:nM_z} and for $\rho_\pm$ by \cref{eq:markovpm,eq:nM_pm}. These results cover the universal dynamics for all times $t > \hbar/\xi$, with a domination of the non-Markovian decay for $t < \tau_T = \hbar/2\pi k_B T$ and a crossover to the standard Markovian exponential decay at $t > \tau_T$. For times $t < \hbar/\xi$, the evolution is non-universal but from an expansion of \cref{eq:master} around $t=0$ for small times we see that the onset of the decay is quadratic in time, $\rho\approx 1-i\Sigma(0)t^2/2$, and hence causes only a very small lowering of the amplitudes before the onset of the universal behavior. The latter starts logarithmically $\sim \ln(\xi t/\hbar)$ for both $\rho_z$ and $\rho_{\pm}$. This is indeed the signature of a Fermi edge singularity many-body reaction \cite{Mahan1,Mahan2,Anderson1,Nozieres} that is triggered by the local spin-spin interaction, and this behavior would turn into Kondo correlations if the interaction could cause an arbitrary number of spin flips \cite{Anderson2,Anderson3}.

In \cref{fig:fulldecay}, we show the full decay of the different components in the cross-over region from non-Markovian to Markovian behavior around $t \sim \tau_T$. Figure \ref{fig:fulldecay}(a) shows the comparison of $\rho_{z}=\rho_{\pm}$ at $b_z^I=0$ with a pure Markovian exponential decay $\rho(t) = e^{-t/T_{1,2}}$. Since the Markovian decay times $T_{1,2}$ are on the order of $\tau_T/\alpha$ (far outside the plotted range) any focus on times around $T_{1,2}$ misses the non-Markovian behavior. However, since the non-Markovian amplitude is on the order of $\alpha$, the amplitude of the Markovian decay is reduced from $1$ to $[1-O(\alpha)]$. The fast non-Markovian decay thus results in a fast initial slip \cite{Geigenmueller,Haake1985,Suarez,Yu,Hell}, a systematic offset of the amplitude of $O(\alpha)$ at all times $t>\tau_T$, which we show at zero field in \cref{fig:fulldecay}(a) and at nonzero field in \cref{fig:fulldecay}(b). The offset is detectable by extrapolating the Markovian behavior back to time $t=0$ [light dashed line in panel (a)]. In panel (c) we use a logarithmic time scale to enhance the visibility of the non-Markovian features at a field $b_z \sim k_B T$ (solid lines), and at a field $b_z \gg k_B T$ (dashed lines) at which the oscillations in $h(t,b_z^I)$ become visible.

\newpage

%
%
%
\section{Manipulating the environment: \\Cooling protocol}\label{sec:Cooling}

The notable feature of the non-Markovian decay is that it is fast and initially always temperature independent. It may thus be possible to put this property to use in an application that depends on the decay of the spin system but is in practice limited by the diverging $T_{1,2}$ times at very low temperatures. Such a situation indeed arises in cooling techniques based on a cold spin system. Adiabatic demagnetization is a standard technique that allows reaching very low temperatures of a spin system such that the latter can be used as a refrigerant for cooling another system \cite{Pobell}. The technique relies on the fact that the entropy of the spin system depends on the ratio of magnetic field and spin temperature, $B/T_I$, such that the adiabatic reduction of an initially strong magnetic field leads to lower spin temperatures $T_I$. But the efficiency as a refrigerant then depends further on the thermal equilibration between the spin system and the rest of the system, including the lattice as well as the electrons. This process relies on the relaxation of the spin system, which is governed by the $T_1$ time. The Korringa relation $T_1 \propto 1/T$ provides the crucial bottleneck in that the increasingly long relaxation times at low temperatures make the cooling ineffective against heat leaks. For instance, nuclear spins in rhodium can be cooled to temperatures below 100 pK but the electron and phonon temperatures are limited to about 0.1 mK, likely due to this effect \cite{Knuuttila,Hakonen}. In other bulk metals such as platinum, electron temperatures of $1.5$ $\mu$K can be reached \cite{Wendler}, but also here a notable discrepancy with the nuclear spin temperature of 0.3 $\mu$K persist. More challenging is the cooling of semiconductors or nanostructures because of their much lower thermal conductivity, and the lowest reached electron temperatures are here in the mK range \cite{Clark,Scheller,Palma1,Palma2,Zyuzin}.

In the following, we thus examine the possibility of using the temperature-independent non-Markovian decay to transport heat out of the electron bath to speed up the slow thermal relaxation process described by the Markovian decay. Quantum thermodynamic cooling protocols explicitly based on nonequilibrium physics have been suggested before (see, e.g., Refs.\ \cite{Haertle,Chen2010,Campo2014,Funo}). We complement these proposals by using our explicit knowledge of the non-Markovian decay to design a protocol with a strong focus on speeding up the cooling process. We should also note that such a cooling protocol goes beyond a thermodynamic cycle because intensive thermodynamic quantities such as the spin temperature cannot be defined during the quantum coherent evolution.
Since the non-Markovian decay has an amplitude of order $\alpha$, this approach needs repeated reinitialization of the spin system. This could be done, for instance, through femtosecond resolved optical pumping of electron spins \cite{Kikkawa} or nuclear spins \cite{Salis2001a,Salis2001b}, by optical pumping of hole spins \cite{Gerardot}, or by partial measurements \cite{Blok}.

\begin{figure}[t!]
    \includegraphics{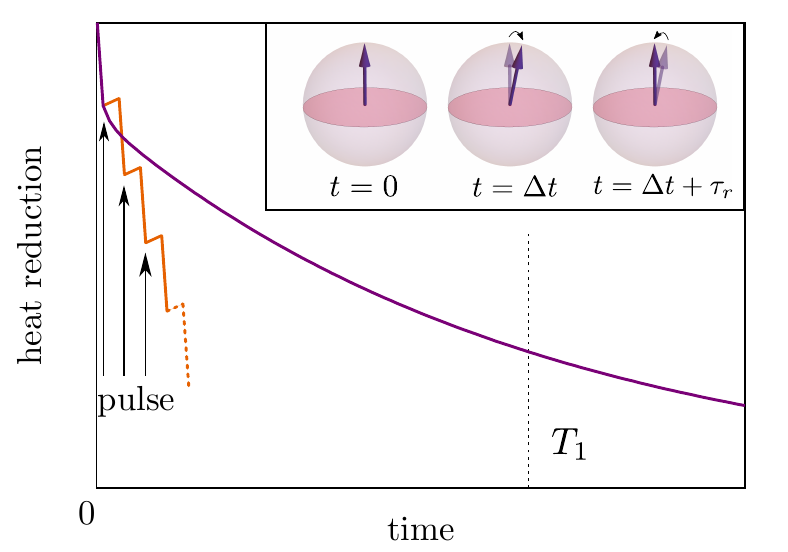}
    \caption{Illustration of the heat transported out of the electronic bath as a function of time. The purple line sketches the conventional relaxation based on the Markovian decay (after the initial slip), where the cooling time scale is set by $T_1$. The orange curve sketches the proposed speed up by a cooling protocol based on a repetition of the initial non-Markovian decay by repeated reinitializations of the spin state. The inset shows the spin state on the Bloch sphere at different stages of one cooling cycle.}
    \label{fig:pumpcycle}
\end{figure}

In \cref{fig:pumpcycle}, this idea is sketched. We show the heat $\Delta Q$ transported out of the electronic system as a function of time. We will see in \cref{eq:heat} below that $\Delta Q$ is proportional to the decay of $\rho_z$. Therefore, $\Delta Q$ has a fast initial non-Markovian decay followed by a slow relaxation set by the relaxation time $T_1$ (dark purple curve). Through a repeated resetting of the spin polarization at the end of the non-Markovian regime, as shown in the inset, we can repeat the fast initial drop of $\Delta Q$ and transport heat out of the system more quickly (bright orange curve).
%
%

The curve shows, of course, an idealized situation and the cooling effect must be complemented by the possible repeat time, the reheating by the pumping, and the influence of external heat leaks. Our goal in the following analysis is to demonstrate under which conditions such a cooling protocol can become effective, and to show that this can be achievable with state-of-the-art techniques.

Instead of a single impurity spin $I$, we require now a macroscopic ensemble of such spins. But if the direct interaction between these spins is weak, we can treat each spin individually and use the results found in the previous sections. Without spin-spin interaction, the impurity spin Hamiltonian $H_I$ becomes
\begin{align}
    H_I = \sum_{i=1}^{N_I} \mu_I g_I B_z I^z_i,
\end{align}
which is the same as the Zeeman term in \cref{eq:H1} but now with the index $i=1,\dots,N_I$ labeling the impurity spins $\mathbf{I}_i$. Under the assumption of independent spins and the assumption that we can neglect the coupling between the impurity spins mediated through the electron system, we can write the reduced density matrix for the spin system as the product $\rho_I = \rho_{I,1}\otimes\dots\otimes\rho_{I,N_I}$. Assuming identical initial conditions, the time evolution of $\rho_I$ consists of $N_I$ copies of the same evolution of a single spin.

The energy current $J_I$ of the spin system can then be defined by
\begin{align}
	J_I = \frac{d}{dt}\,\text{Tr}_\text{el}[H_I\rho],
\end{align}
such that $J_I>0$ corresponds to an energy flow from the environment into the spin system.
The heat $\Delta Q(t)$ transferred up to a time $t$ is given by
\begin{align}
    \Delta Q(t) &= \int_0^{t} dt'J_I(t')
    =N_I\mu_ng_IB_zI\left[\rho_z(t)-\rho_z(0)\right],
    \label{eq:heat}
\end{align}
where $\rho_z(t)$ is the decay of a single representative spin and the factor $N_I$ takes the ensemble into account.
To focus on the non-Markovian decay in $\Delta Q$, we consider time scales $t\ll\tau_T$ at which, from \cref{eq:nM_z}, the dynamics of $\rho_z(t)$ becomes temperature independent and leads to
\begin{align}
    &\Delta Q(t)
    =-4 \alpha Q_z \rho_0 \ln\left(\frac{\xi t}{\hbar}\right),
\end{align}
with $Q_z=N_I\mu_n g_I B_z I = N_I b_z^I$ and $\rho_0 = [\rho_z(0)-\rho_z^\text{eq}]$.

The spins absorb energy if $\Delta Q < 0$ and therefore the spins should initially minimize their Zeeman energy such that $B_z [\rho_z(0)-\rho_z^\text{eq}]<0$. For $B_z>0$, this means we should choose $\rho_z(0) \approx -1$, corresponding indeed to the ground state of ultracold spins, instead of the highly excited $\rho_z(0) = +1$ chosen for illustration in the previous sections. We then obtain $\rho_0 = - [1-\tanh(b_z^I/2\pi k_B T)] = - |\rho_0|$. For an optimal cooling, the amplitude $B_z |\rho_0|$ should be as large as possible, but large $B_z$ lead to a small $|\rho_0|$. Since $|\rho_0|$ depends on the ratio $b_z^I/2\pi k_B T$, the optimum is obtained by tuning $b_z^I$ to the order of $2\pi k_B T$ and consequently $B_z |\rho_0| \sim T$.

We now want to consider a cooling cycle where we utilize the temperature-independent non-Markovian decay to transfer heat from the electronic bath into the impurity spin system. The relevant time scale for the duration of one cycle is $\Delta t + \tau_r$, where $\Delta t \ll T_1$ is the time interval of the spin decay in the fast non-Markovian regime and $\tau_r$ is the time needed to reinitialize the impurity spin. In the inset of \cref{fig:pumpcycle}, a sketch of the cooling cycle and the corresponding state of the impurity spin on the Bloch sphere is shown. In the short time $\Delta t$, the spin decays by a small amplitude proportional to $\alpha$, according to \cref{eq:nM_z}. After the initial slip, the impurity spin needs to be projected back onto its initial state, during which the system is necessarily reheated by an amount $Q_r$.

The total heat $Q_\text{el}$ transferred out of the electron system over $N_P$ cooling cycles can then be written as
\begin{align}
    Q_\text{el}(t)=&N_P\left[-\Delta Q(\Delta t)+ Q_r\right]+t J_\text{ext}^Q, \label{eq:Qel}
\end{align}
where $-\Delta Q(\Delta t)$ is the non-Markovian cooling per cycle, $Q_r$ is the heat deposited during each reinitialization process, and $t J^Q_\text{ext}$ takes into account external heat leaks generating a continuous inflowing heat current $J^Q_\text{ext}$. The total time $t$ of the process is $N_P (\Delta t + \tau_r)$.

Cooling is possible if $Q_\text{el}<0$. By dividing \cref{eq:Qel} by $N_P |4\alpha Q_z \rho_0|$ and absorbing $Q_r/|4\alpha Q_z \rho_0|$ in the logarithm, we can write this condition as
\begin{align}
    q_\text{ext} < q_\text{cp},
    \label{eq:constraint}
\end{align}
where
\begin{equation}
	q_\text{cp}
	= \frac{\tau_0}{\Delta t + \tau_r} \ln\left(\frac{\Delta t}{\tau_0}\right)
\end{equation}
is a dimensionless quantity measuring the efficiency of the cooling protocol,
\begin{equation}
	q_\text{ext} = \frac{\tau_0 J^Q_\text{ext}}{|4 \alpha Q_z \rho_0|}
\end{equation}
measures the influence of the external heat leaks, and
\begin{equation} \label{eq:tau_0}
	\tau_0 =\hbar\xi^{-1} \exp(Q_r/\lvert4\alpha Q_z \rho_0 \rvert)
\end{equation}
sets the characteristic time for the protocol.
The time $\tau_0$ should fulfill $\tau_0 < \tau_T$ as only then the non-Markovian decay is effective.
This provides a condition on $Q_r$,
\begin{equation} \label{eq:constraint_Qr}
	Q_r < |4 \alpha Q_z \rho_0| \log(\xi \tau_T/\hbar).
\end{equation}
Since $Q_z \rho_0$ should be chosen proportional to $T$, this condition puts a not surprising constraint on the lowest reachable temperatures set by the heating by reinitialization. Since the dependence of the protocol on the reinitialization time $\tau_r$ is less important, an optimization through the interaction time $\tau_r$ should be possible. Notice also that a tuneability exists through the electron density as $\xi \sim E_F$.

\begin{figure}[t!]
    \includegraphics{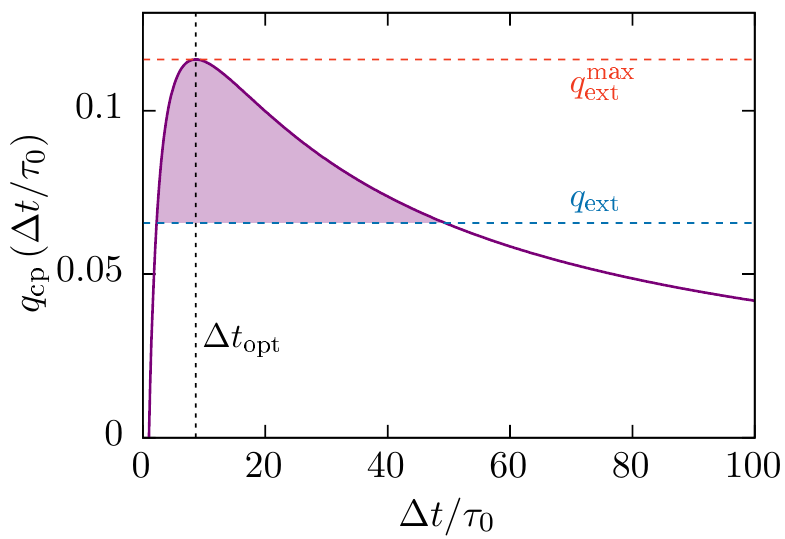}
    \caption{Plot of the condition imposed by \cref{eq:constraint} as a function of $\Delta t/\tau_0$. The curve shows $q_\text{cp}$ for $\tau_r/\tau_0=10$ in comparison with an arbitrarily chosen external heat leak $q_\text{ext}$. The shaded area marks the region where $q_\text{cp}>q_\text{ext}$ and cooling is possible. The orange dashed line marks the maximum value $q_\text{ext}^{\text{max}}$ for an external heat leak. The black dashed line marks the optimal time $\Delta t_\text{opt}$ for the most efficient heat transport. The $T_1$ time in this type of graph should generally lie at large time far outside the plotted range.}
    \label{fig:cooling_noT}
\end{figure}

The influence of the external heat leaks is assessed through \cref{eq:constraint}. Figure \ref{fig:cooling_noT} shows $q_\text{cp}$ as a function of $\Delta t/\tau_0$ in comparison with a choice of $q_\text{ext}$. The shaded area marks the parameter region where \cref{eq:constraint} is fulfilled and cooling the system with the pumping protocol is possible. There is naturally a maximum heat leak $q_\text{ext}=q_\text{ext}^\text{max}$ beyond which cooling is no longer possible, indicated by the maximum of the curve.
Maximizing the amount of heat that can be carried out of the electronic system throughout the whole cooling process leads to an optimal time $\Delta t$ between pulses
\begin{align}
    \Delta t_\text{opt}=\tau_0 e^{1+W(\tau_r/e \tau_0)} = \tau_r/W(\tau_r/e\tau_0),
\end{align}
where $e$ is Euler's number and $W(z)$ is the Lambert $W$ function, defined as the inverse of the function $z(W) = W \exp(W)$. If $z=\tau_r/ e \tau_0 \ll 1$, large heating by the pumping is implied and $W(z)\approx z/e$, i.e., $\Delta t \sim \tau_0$. In the opposite limit $z=\tau_r/ e \tau_0 \gg 1$, we have $W(z)\approx \ln(z)$, which leads to $\Delta t \sim \tau_r$. Upon the constraint of \cref{eq:constraint_Qr}, we see that $\Delta t_\text{opt}$ is thus within the range set by $\tau_r$ and $\tau_T$.
%

Taking \cref{eq:Qel}, we can estimate how many pump cycles are possible by looking at the temperature of the electronic system $T_\text{el}(t)$, assuming an equilibrium state of the bath itself,
\begin{align}
    T_\text{el}(t) &= T_\text{el}(0) +\frac{Q_\text{el}(t)}{C_\text{el}},
\end{align}
where $C_{\text{el}}$ is the specific heat for the electronic system. Before we start pumping the system, we can assign an initial temperature $T_I(0)$ to the impurity spins. Although through the pumping and reinitializations the concept of a spin temperature $T_I$ becomes no longer meaningful, we shall use $T_I(0)$ as a lower bound, but due to the temperature dependence of $q_\text{cp}$ and $\tau_0$ it may be an optimistic bound for the final electronic temperature. Setting thus the minimum temperature the electrons can reach to  $T_\text{el}(t)=T_I(0)$ and assuming that the temperature dependence of the parameters is otherwise weak, we are led to
\begin{align}
    N_P = \frac{C_\text{el}[T_\text{el}(0) - T_I(0)]}{Q(\Delta t) - Q_\text{el}^P - \left(\Delta t+\tau_r\right) J_\text{ext}^Q}.
\end{align}
During each cycle, a small amount of heat proportional to $\alpha$ is transferred from the electrons into the impurity spin system via spin flips. Thus, the efficiency of the cooling process relies on a fast repetition such that $N_P \sim 1/\alpha$ repetitions can be made faster than the $T_1$ time.

The small parameter $\alpha = (\nu_0 A)^2$ is set by the ratio of the coupling constant $A$ and the Fermi energy, but $\nu_0$ contains
also the ratio $n_e/n_I$ of electron density $n_e$ over impurity spin density $n_I$ \cite{Simon_1,Simon_2,Braunecker_1,Braunecker_2}.
The parameter is thus highly dependent on the considered system and can vary from very small $\alpha \sim 10^{-10}$, as is characteristic for nuclear spins in bulk metals \cite{Pobell},
to $\alpha \sim 10^{-4}$ for paramagnetic spins in correlated metals \cite{Bonville}, and to $\alpha > 1$ in magnetic semiconductors \cite{Okabayashi,Jungwirth}, which is, however, beyond the validity of our approach.

As an example, let us consider a semiconductor with nuclear spins, e.g., GaAs, with $A=90$ $\mu$eV and $E_F$ on the order of meV, such that $\alpha \sim 10^{-8}$.
For such small $\alpha$, a direct observation of the non-Markovian behavior is elusive but it provides a good illustration that the
cooling protocol could even then become useful.
The corresponding $T_1 = \hbar/4\alpha\pi k_B T$ is at $T \sim 0.1$ mK on the order of seconds. The suggested heat transfer protocol becomes effective if we can repeat it $N_P \sim 1/\alpha$ times within the $T_1$ time. Thus, the maximum time $\Delta t+\tau_r$ between two pulses should be limited by  $\Delta t+\tau_r < \alpha T_1 \sim 10$ ns, which means that both $\Delta t$ and $\tau_r$ can be on the order of 10 ns.

Recall that $\tau_0$ in \cref{eq:tau_0} is given by $\hbar/\xi \sim \hbar/E_F$, which is in the picosecond range, times an exponential. The latter is then allowed to grow to $10^3$-$10^4$ such that the requirement on the exponent $Q_r/|4 \alpha Q_z \rho_0|$ is not too stringent. Furthermore, a reinitialization time $\tau_r \sim$ 10 ns is short but not beyond the reach of modern experimental techniques. We must also recall that it corresponds to a reinitialization of a tiny decay of amplitude $\alpha$ such that the limiting factor will be given by the electronics for the repetition rate rather than the physical manipulation of the spin.

Moreover, tuning $\alpha$ to larger values works in favour of this cooling protocol. Instead of influencing $\alpha$ through $E_F$, we can also substantially enhance it by considering different materials as outlined above. Values of $\alpha \sim 1$ can become rather common, but we must emphasize that our theory depends on the smallness of $\alpha$ and that at the resulting small $T_1$ times the standard cooling techniques may remain more effective.

%
%
\section{Conclusions}

In this work, we discussed a unifying analytical approach to nonequilibrium quantum thermodynamics, coherent non-Markovian evolution, and backaction from the system on the bath for quantum spins embedded in an electron conductor. We addressed three goals in particular:
(\emph{i}) the presentation of a general framework allowing us to obtain the full dynamics of a spin system in a bath with non-Markovian memory effects; (\emph{ii}) the application when the bath is an itinerant fermionic system; and (\emph{iii}) the proposition of a cooling protocol based on the temperature insensitive non-Markovian quantum fluctuations. The development of the framework was based on a careful expansion of the generalized master equation for the spin's density matrix and on an extraction of the physical consequences from the pole structure in the complex Laplace space. The memory effect arising from the backaction of the fermionic system on the spin dynamics is indeed governed by the coherent excitations of the fermionic bath, which each are represented by one of the poles. The collective of the poles then defines the non-Markovian dynamics, and we have presented a systematic method for accessing the dynamics. This method is limited to a small coupling strength between the spin and the bath degrees of freedom, encoded in the parameter $\alpha$ defined in \cref{eq:alpha}. But the method is not limited to the considered fermionic system, and an extension of this work to strongly correlated systems will be of special interest. In the latter systems, the internal interactions lead to strong time correlations and therefore to an enhanced non-Markovian contribution to the spin dynamics. An analysis of the memory effects will thus provide a time-resolved access to the strong correlation physics.
Furthermore, the method is not restricted to fermionic environments. As long as the bath correlators can be computed, a similar analysis of the pole structure can follow for a bosonic environment as well.

As an archetype for a correlated response, we have considered the example of a free itinerant electron gas. Such an electron gas resembles a critical correlated system in the sense that its quantum and thermal fluctuations are governed by a small number of parameters, here the electron temperature $T$ and the Fermi energy $E_F$. For such an electronic bath, we have provided an explicit solution to the full free induction decay of the spin system, crossing over from the initial fast non-Markovian decay to the conventional Markovian exponential decay. The latter is characterized by the times $T_1$ and $T_2$, whereas the non-Markovian decay occurs on the much shorter electronic thermal time and remains detectable at longer times through the initial slip, the systematic offset of the Markovian decay from the expected initial value at time zero.

At short times, the non-Markovian decay is temperature independent and determined entirely by quantum fluctuations. Since this decay can be tuned to transport heat from the electron to the spin system, we have finally proposed a cooling protocol based on a repeated triggering of the non-Markovian decay by reinitializations. Such a method could overcome the diverging time scale $T_1 \propto 1/\alpha T$ that limits the efficiency of adiabatic demagnetization cooling. Although the method is probably not suitable to concurrence with the standard cooling methods for bulk metals, we have provided an estimate that it could become effective in cooling semiconductor structures. For semiconductors, the $T_1$ times are much longer and are responsible for the bottleneck that has prevented experiments from reaching electron temperatures below a millikelvin so far.

\begin{acknowledgments}
We thank S.~Parameswaran, R.~Whitney, F.~Pollock and T.~Stace for useful discussions. S.M.\ acknowledges the support by the EPSRC under Grant No.\ EP/L015110/1. D.L.\ and J.K.\ acknowledge the support by the Swiss National Science Foundation and NCCR QSIT. The work presented in this paper is theoretical. No data has been produced and supporting research data is not required.

\end{acknowledgments}

%
\appendix
%
\section{Master equation for reduced density matrix} \label{app:master}
To gain access to the temporal correlations or memory effects, we use the projection operator method \cite{Breuer,Sauermann}. In the following, we provide a short derivation of \cref{eq:master} in the main text. The starting point is the Nakajima-Zwanzig equation, a generalized exact master equation for the density matrix $\rho(t)$
\begin{align}
    \frac{d}{dt} P\rho(t)=-iPLP\rho(t)- i \int_0^t dt'\,\Sigma(t-t')\rho(t'),
    \label{eq:eom}
\end{align}
with the memory kernel  $\Sigma (t-t')$ .
The superoperator $L$ is the Liouvillian defined by the Hamiltonian $H$, i.e., $LO=~[H,O]$ for any operator $O$. The memory kernel,
\begin{equation}
	\Sigma(t-t')= - iPLQ e^{-iQL(t-t')} LP,
\end{equation}
captures the whole history of the system up to time $t$. The projection operators $P$ and $Q$ obey $P+Q=~\mathbb{1}$, $P^2=P$, $Q^2=Q$, and $PQ=0$.
We choose the projection operator $P$ as $PO=~\rho_\text{el}\otimes~\text{Tr}_\text{el}\left[O\right]$, with $\rho_\text{el}$ being the equilibrium density matrix of the bath.
 The trace over the bath degrees of freedom $\text{Tr}_\text{el}[\cdot]$ defines the reduced density matrix describing the spin system $\rho_I=\text{Tr}_\text{el}\left[\rho\right]$.
Splitting the Hamiltonian into  $H=H_0+H_\text{int}$, we can write the Liouvillian $L = L_0 + L_\text{int}$ accordingly. Then, the first term in \cref{eq:eom} including $L_0$, which describes the eigendynamics of the spin system, drops out due to the choice of the rotating frame of reference. The part containing $PL_\text{int}P$ vanishes since the system is spin conserving and does not acquire a net magnetic moment. Using the definition of the projector $P$ leads to the master equation, \cref{eq:master}, solved in the main text for the reduced density matrix $\rho_I$
\begin{align}
    \frac{d}{dt}\rho_I(t)=- i \int_0^t dt'\,\Sigma_I\left(t-t'\right)\rho_I(t'),
    \label{eq:master_app}
\end{align}
with the reduced memory kernel
\begin{align}
    \Sigma_I(t-t')=- i \text{Tr}_\text{el}\left[Le^{-iQL(t-t')}QL_\text{int}\rho_\text{el}\right].
    \label{eq:memory_app}
\end{align}
To solve the master equation in \cref{eq:master_app} for the full time evolution we analyze it in Laplace space. The Laplace transform of the generalized master equation, $f(t)\to~\tilde{f}(s)=~\int_0^\infty dt\,e^{-ts}f(t)$, with $\text{Re}(s) > 0$, is given by
\begin{align}
    \tilde{\rho}_I(s) &=\left[s\mathbb{1}+i \tilde{\Sigma}_I(s)\right]^{-1}\rho_I(t=0),
\end{align}
with the identity operator $\mathbb{1}$, and the Laplace transform of the reduced memory kernel $\tilde{\Sigma}_I(s)$ is
\begin{align}
    \tilde{\Sigma}_I(s)=-i\text{Tr}_\text{el}\left[L\left(s\mathbb{1}+iQL\right)^{-1}QL_\text{int}\rho_\text{el}\right].
\end{align}
The operator $\left(s\mathbb{1} + i LQ\right)^{-1}$ obeys a Schwinger-Dyson equation and iteration of this identity provides an expansion in powers of the interacting part $L_\text{int}$ of the full Liouvillian \cite{Sauermann,LossCoish}
\begin{align}
    &\left(s\mathbb{1} + iQL\right)^{-1} =\notag\\
	&\ \ \ \ \sum_{n=0}^\infty \left(s\mathbb{1} + QL_0\right)^{-1}\left[-iQL_\text{int}\left(s\mathbb{1} + i Q L_0\right)^{-1}\right]^n.
    \label{eq:series_memory}
\end{align}
For odd $n$, the corresponding term in the memory kernel $\tilde{\Sigma}_I(s)$ is zero since the Hamiltonian $H_0$ is spin conserving. In the main text, we used the Born approximation up to second order in the interaction. This corresponds to the first term in the series expansion in \cref{eq:series_memory}.

\section{Expression for the memory kernel $\tilde{\Sigma}(s)$}
\label{app:memory}
Assuming spin-1/2 for the impurity spin, every operator $O$ can be decomposed in the basis of the spin system spanned by  $\{I_\uparrow,I_\downarrow,I_-,I_+\}$. The operator $O$ can then be written as
\begin{align}
    O=o_0\mathbb{1}+o_\uparrow I_\uparrow+o_\downarrow I_\downarrow + o_+ I_- + o_-I_+,
\end{align}
Using the impurity spin basis in the rotating frame of reference, the superoperators, $[L_0],\ [L_\text{int}],\ [\tilde{\Sigma}(s)]$, can be expressed as a $4\times4$ matrix. Here, we denote the matrix representation of the superoperators with $[\cdot]$. To find expressions for the superoperators, we use lower case operators which act only on the bath. The Hamiltonian for the bath can then be written as $H_0=h_0\otimes\mathbb{1}_I$. For the interacting part, we decompose the spin-spin interaction
\begin{align}
    H_\text{int}=h_\uparrow I_\uparrow+h_\downarrow I_\downarrow + h_+ I_- + h_-I_+.
\end{align}
In the case of the spin-spin interaction discussed here, $H_\text{int}=~A\mathbf{S}\cdot\mathbf{I}$ the bath operators are
$h_\uparrow=-h_\downarrow=\pm A S_z/2=h_z$ and $h_\pm=2AS_\pm$.
Within the basis of the spin-1/2, $\{I_\uparrow,I_\downarrow,I_-,I_+\}$, the interaction part of the Liouvillian $[L_\text{int}]$ is given by
\begin{align}
[L_\text{int}]=
    \begin{pmatrix}
        L^-_{h_z}&0&h^L_-&-h^R_+\\
        0&-L^-_{h_z}&-h^R_-&h^L_+\\
        h^L_+&-h^R_+&-L^+_{h_z}&0\\
        -h^R_-&h^L_-&0&L^+_{h_z}
    \end{pmatrix}.
\end{align}
Here, superscript $-$ refers to the commutator, $L_{h_z}^-o =~ [h_z,o]$, and the superscript $+$ refers to the anticommutator,\,$L_{h_z}^+o=~\{h_z,o\}$. $L,\ R$ denote if the operator acts from the left- or right-hand side.
Finally, the memory kernel within the spin system's basis can be expressed as
\begin{align}
    \left[\tilde{\Sigma}_I(s)\right]=
    \begin{pmatrix}
            F_1&-F_2&0&0\\
            -F_1&F_2&0&0\\
            0&0&F_-+F_z&0\\
            0&0&0&F_++F_z
    \end{pmatrix}.
\end{align}
The entries of the memory kernel $F_{1,2,z,\pm}(s)$ with $\text{Re}(s)>0$ are the Laplace transforms of spin-spin correlation functions
\begin{align}
    F_1(s)&=\frac{2A^2}{i}\int_0^\infty dt \,e^{-st}\,\text{Re}\left[\langle S_{-,0}(t)S_{+,0}(0)\rangle\right],\label{eq:cor1}\\
    F_2(s)&=\frac{2A^2}{i}\int_0^\infty dt \,e^{-st}\,\text{Re}\left[\langle S_{+,0}(t)S_{-,0}(0)\rangle\right],\label{eq:cor2}\\
    F_-(s)&=\frac{A^2}{i}\int_0^\infty dt \,e^{-st}\langle\left\{ S_{+,0}(t),S_{-,0}(0)\right\}\rangle,\label{eq:cor3}\\
    F_+(s)&=\frac{A^2}{i}\int_0^\infty dt \,e^{-st}\langle\left\{ S_{-,0}(t),S_{+,0}(0)\right\}\rangle,\label{eq:cor4}\\
    F_z(s)&=\frac{A^2}{2i}\int_0^\infty dt \,e^{-st}\langle\left\{ S_{z,0}(t),S_{z,0}(0)\right\}\rangle,\label{eq:cor5}
\end{align}
for $t\geq0$. For a system with $SU(2)$, symmetry all these expressions are equal, and an example for such a situation is the simple Fermi gas considered in the main text. But it should be emphasized that the $SU(2)$ symmetry of the Fermi gas holds only in the laboratory frame since the transformation to the rotating frame \cref{eq:rot_frame} explicitly breaks it. For the evaluation of the fermionic correlators, it is therefore important to go back to the laboratory frame, in which the latter equations become
\begin{align}
    F_1(s)&=\frac{2A^2}{i}\int_0^\infty dt\,e^{-st} \text{Re}\left[e^{ i b_z^I t}\langle S_{-,0}(t)S_{+,0}(0)\rangle_\text{lab}\right],\label{eq:lab1}\\
    F_2(s)&=\frac{2A^2}{i}\int_0^\infty dt\,e^{-st} \text{Re}\left[e^{-i b_z^I t}\langle S_{+,0}(t)S_{-,0}(0)\rangle_\text{lab}\right],\label{eq:lab2}\\
    F_-(s)&=\frac{A^2}{i}\int_0^\infty dt \,e^{-(s+ib_z^I)t}\langle\left\{ S_{+,0}(t),S_{-,0}(0)\right\}\rangle_\text{lab},\label{eq:lab3}\\
    F_+(s)&=\frac{A^2}{i}\int_0^\infty dt \,e^{-(s-ib_z^I)t}\langle\left\{ S_{-,0}(t),S_{+,0}(0)\right\}\rangle_\text{lab},\label{eq:lab4}\\
    F_z(s)&=\frac{A^2}{2i}\int_0^\infty dt \,e^{-st}\langle\left\{ S_{z,0}(t),S_{z,0}(0)\right\}\rangle_\text{lab},\label{eq:lab5}
\end{align}
Here, $b_z^I$ is the prefactor for the Zeeman term of the impurity spin in \cref{eq:H1}. Even with the underlying $SU(2)$ symmetry, these functions are generally distinct, and only at $b_z^I=0$ will the $SU(2)$ symmetry of the Fermi gas cause $F_1(s) =~ F_2(s) = ~F_\pm(s) =~F_z(s)$.

\section{Spin-spin correlator for a Fermi gas} \label{app:correlator}

To calculate the Laplace transform of the spin-spin correlation functions \cref{eq:cor1,eq:cor2,eq:cor3,eq:cor4,eq:cor5} we rewrite each spin $\mathbf{S}$ as
\begin{align}
    \mathbf{S} = \sum_{kk'\sigma\sigma'} c^\dagger_{k\sigma}\boldsymbol{\tau}_{\sigma,\sigma'} c_{k'\sigma'},
\end{align}
where $c_{k\sigma}$ are the electron operators and $\boldsymbol{\tau} = ({\tau^x,\tau^y,\tau^z})$ is the vector of Pauli matrices.
Let us focus first on the zero-field case, $b_z^I=0$, in which from the $SU(2)$ symmetry of the Fermi gas it follows that $F_1(s) =~ F_2(s) = ~F_\pm(s) =~F_z(s)$.
This allows us to restrict the calculation to $F_z(s)$, which for the Fermi gas is given by
\begin{align}
    &F_z(s) = \frac{A^2}{2 i}\frac{a^{2d}}{(2\pi)^d}\int_0^\infty dt\,e^{-ts}\notag\\
    &\times\sum_{kk'\sigma}\Big[e^{i(\epsilon_k-\epsilon_{k'})t}\langle c^\dagger_{k\sigma} c_{k\sigma}\rangle\langle c_{k'\sigma} c^{\dagger}_{k'\sigma}\rangle + \text{c.c.}\Big].
    \label{eq:fz}
\end{align}
In this expression, the sum runs over all $k,k'$ within the first Brioullin zone with the $d$-dimensional unit cell volume $a^{2d}$, and $\sigma$ is the spin index.
To evaluate the $k$ summations, we introduce the density of states $\nu(\epsilon)=~a^d\lvert d^dk/d\epsilon\rvert/(2\pi)^d$ and integrate over $\epsilon$ instead. To capture the important physics about the Fermi energy $E_F$ without being troubled by the nonuniversal high-energy contributions near the band edges, we write the density of states as $\nu(\epsilon)=\nu_0\exp{\left(-\lvert \epsilon\rvert/\xi_0\right)}$, where we choose to set the zero of $\epsilon$ to the Fermi energy and where $\xi_0 \sim E_F$ is a high-energy cutoff. This approximation maintains a constant $\nu(\epsilon) = \nu_0$ near the Fermi surface but the exponential provides a finite bandwidth $\xi_0$ without introducing high-energy artifacts. As a consequence, all nonuniversal (very short time) behavior that depends on the structure of the entire band will be absorbed in the $\xi_0$, and the $\xi_0$-independent part of the response function represents the universal low-energy physics. Further corrections of order $\xi_0^{-1}$ will be neglected. It is then convenient to introduce the function
\begin{align}
    F(s)
	&= -\frac{1}{\nu_0^2} \int_0^\infty dt\,e^{-ts}\int d\epsilon \, \nu(\epsilon) e^{i \epsilon t} f(\epsilon)
	\notag\\
	&\times \int d\epsilon' \, \nu(\epsilon') e^{-i \epsilon' t} \bigl[1-f(\epsilon')\bigr]
	\notag\\
	&= - \int_0^\infty dt \, e^{-ts} \left[ \int d\epsilon \, e^{-|\epsilon|/\xi_0} e^{i \epsilon t} f(\epsilon)\right]^2,
	\label{eq:F_int}
\end{align}
where $f(\epsilon)$ is the Fermi-Dirac distribution. For the last identity, we have used that when $\epsilon$ is measured off the Fermi surface that $\nu(\epsilon) = \nu(-\epsilon)$ and $f(-\epsilon) = 1-f(\epsilon)$.
With $f(\epsilon) = [1+\exp(\epsilon/k_B T)]^{-1}$, the energy integration leads to
\begin{align}
    F(s)&= - \int_0^\infty dt\,e^{-ts}\frac{\left(\pi k_B T\right)^2}{\sin^2{\left(\pi k_B T (it+\xi_0^{-1})\right)}}.
    \label{eq:laplace}
\end{align}
The cutoff $\xi_0$ regularizes the short time divergence of the remaining time integral associated with the nonuniversal high-energy sector, and the $\xi_0$-independent contribution encodes the universal low energy physics.
As mentioned, we neglect in the remaining time integral all orders $\xi_0^{-1}$ or smaller. We then obtain
\begin{align}
        F(s)=&\ i\xi_0-\pi k_B T  \notag\\
        &+s\left[\ln{\left(\frac{2\pi k_B T}{i \xi}\right)}+\psi\left(1+\frac{s}{2\pi k_B T}\right)\right],
        \label{eq:F(s)}
\end{align}
where $\psi(z)$ is the digamma function, and $\xi$ is a slight renormalization of $\xi_0$ absorbing the constant $\psi(1)/2$ in the logarithm.
For the response function $F_z$, this $F$ contributes the first term of the commutator of \cref{eq:cor5}. The second term has a reverse time evolution in the expectation values, leading to $e^{-i \epsilon t}$ instead of $e^{i \epsilon t}$ in \cref{eq:F_int}. As a result, the integral evaluates to the same $F(s)$ as in \cref{eq:F(s)} but with a conjugation in the cutoff $i \xi_0 \to -i \xi_0$. To avoid any ambiguity, we define the function
\begin{align}
	G(s) &= F(s)|_{i \xi_0 \to -i \xi_0} = -i\xi_0-\pi k_B T  \notag\\
        &+s\left[\ln{\left(\frac{2\pi k_B T}{-i \xi}\right)}+\psi\left(1+\frac{s}{2\pi k_B T}\right)\right].
        \label{eq:G(s)}
\end{align}
In terms of $F$ and $G$, we then have
\begin{equation} \label{eq:F_z_F_G}
	F_z(s) = i \alpha \bigl[ F(s) + G(s) \bigr],
\end{equation}
where $\alpha = (\nu_0 A)^2$ is the small dimensionless coupling parameter.

Notice that all correlation functions are of the form $F+G$ with various arguments (see below), and therefore the nonuniversal terms $\pm i \xi_0$ always cancel and the only dependence on the cutoff remains in the logarithms. This allows us to drop the $\pm i \xi_0$ and use henceforth
\begin{align}
        F(s)&= -\pi k_B T  \notag\\
        &+s\left[\ln{\left(\frac{2\pi k_B T}{i \xi}\right)}+\psi\left(1+\frac{s}{2\pi k_B T}\right)\right],
        \label{eq:f}
\\
	G(s) &= -\pi k_B T  \notag\\
        &+s\left[\ln{\left(\frac{2\pi k_B T}{-i \xi}\right)}+\psi\left(1+\frac{s}{2\pi k_B T}\right)\right],
        \label{eq:g}
\end{align}
Since for the free Fermi gas at zero magnetic field $F_z =~ F_1 =~ F_2 =~ F_\pm$, \cref{eq:F_z_F_G} provides all spin-spin correlators.
For a finite field $b_z^I \neq 0$, we have seen through \cref{eq:lab1,eq:lab2,eq:lab3,eq:lab4,eq:lab5} that we can still express the correlators in terms of the $SU(2)$ symmetric laboratory frame, which means in terms of $F(s)$ and $G(s)$ but with arguments $s$ that are shifted by $\pm ib_z^I$ as follows:
\begin{align}
    F_1(s) &= i\alpha \left[F\left(s-ib_z^I\right) + G\left(s+ib_z^I\right)\right],\label{eq:app1}\\
    F_2(s) &= i\alpha \left[F\left(s+ib_z^I\right) + G\left(s-ib_z^I\right)\right],\label{eq:app2}\\
    F_-(s) &= i\alpha \left[F\left(s+ib_z^I\right) + G\left(s+ib_z^I\right)\right] \notag\\
    &=  F_z\left(s+ib_z^I\right),\label{eq:app3}\\
    F_+(s) &= i\alpha \left[F\left(s-ib_z^I\right) + G\left(s-ib_z^I\right)\right] \notag\\
	&=  F_z\left(s-ib_z^I\right),\label{eq:app4}\\
    F_z(s) &= i\alpha \left[F(s) + G(s)\right].\label{eq:app5}
\end{align}
Notice that only $F_z$ remains unchanged from the zero-field expression.
Notice furthermore that because of the different cutoffs $\pm i \xi$ in $F$ and $G$ the two correlators $F_1(s)$ and $F_2(s)$ are different. Because of this, we obtain the correct equilibrium magnetization $\rho^\text{eq}_z$ derived in \cref{eq:rho_z_eq}.

%
%
\section{Markov poles} \label{app:Markov}
The full time evolution for the reduced density matrix is given by \cref{eq:rhoz,eq:rhopm}, which are evaluated through the sum over the residues at the poles of the integrands. One of the poles of $\rho_z$ is always trivial, $s=0$, and leads to the long time equilibrium value.
The remaining poles can be split in one Markov pole and an infinite number of non-Markovian poles.
In this appendix we evaluate the Markov pole for the different components of the reduced density matrix.

For the longitudinal component $\rho_z(s)$, we split off the $s=0$ pole, allowing us to write
\begin{align}
    \rho_z(s) = \frac{1}{s}\frac{F_2(0)-F_1(0)}{F_1(0)+F_2(0)}+\frac{\rho_0-\frac{F_2(0)-F_1(0)}{F_1(0)+F_2(0)}}{s+iF_1(s)+iF_2(s)}.
\end{align}
The first term produces in \cref{eq:rhoz} a time-independent contribution and hence the equilibrium value $\rho_z^\text{eq}$ obtained in the long time limit $t\to \infty$. Using the results of \cref{app:correlator}, we obtain
\begin{align} \label{eq:rho_z_eq}
	\rho_z^\text{eq} = \frac{F_2(0)-F_1(0)}{F_1(0)+F_2(0)} = -\tanh\left(\frac{b_z^I}{2k_B T}\right),
\end{align}
corresponding indeed to the conventional magnetization of a magnetic moment in a magnetic field.

The second pole of $\rho_z$ responsible for the dynamics is located at
\begin{align}
    s_{z}^\text{M}=&-4\alpha\pi k_B T\notag\\
    &+2\alpha i b_z^I\left[\psi\left(1+\frac{ib_z^I}{2\pi k_B T}\right)-\psi\left(1-\frac{ib_z^I}{2\pi k_B T}\right)\right]\notag\\
    =&-2\alpha \frac{b_z^I}{\tanh\left(\frac{b_z^I}{2 k_B T}\right)},
\end{align}
where for the second equality we have used the identity $2 \text{Im}\psi(1+i x) = -1/x + \pi/\tanh(\pi x)$ \cite[Sec. 5.4.18]{DLMF}.
For the transverse components $\rho_\pm(t)$, the Markov poles sit at
\begin{align}
	&s_{\pm}^\text{M} = -4\alpha\pi k_B T\notag\\
	& \mp i2\alpha b_z^I\left[2\ln{\left(\frac{2\pi k_B T}{\xi'}\right)}+\psi\left(1+\frac{s\mp i b_z^I}{2\pi k_B T}\right)\right],
	\label{eq:mpoles}
\end{align}
where $\xi'$ is a slightly shifted cutoff that absorbs a constant $\psi(1)$ contribution.
The residues for these poles $s_{z,\pm}^\text{M}$ are
\begin{align}
    &\text{Res}\left(s_{z}^\text{M}\right)=\frac{e^{s_{z}^\text{M}t}}{1-4\alpha\left\{\ln{\left(\frac{2\pi k_B T}{\xi}\right)}+\text{Re}\left[\psi\left(1+ \frac{ib_z^I}{2\pi k_B T}\right)\right]\right\}},\\
    &\text{Res}\left(s_{\pm}^\text{M}\right)=\frac{e^{s_{\pm}^\text{M}t}}{1-2\alpha\left[2\ln\left(\frac{2\pi k_B T}{\xi'}\right)+\psi\left(1\mp\frac{i b_z^I}{2\pi k_B T}\right)\right]}.
\end{align}
The real part of the pole $s_{z,\pm}^\text{M}$ defines the relaxation time $T_1$ for the longitudinal component $\rho_z$ of \cref{eq:rhoz} and the decoherence time $T_2$ for the transverse component $\rho_\pm$ of \cref{eq:rhopm} for the Markovian part of the decay (reintroducing $\hbar$ in the following expressions):
\begin{align}
    T_1=&\frac{-1}{s_{z}^\text{M}}=\frac{\hbar}{2\alpha \pi b_z^I}\tanh\left(\frac{b_z^I}{2 k_B T}\right),\\
    T_2=&\frac{-1}{\text{Re}\left[s_{\pm}^\text{M}\right]}
    = \frac{\hbar \tanh\left(\frac{b_z^I}{2 k_B T}\right)}{2\alpha\pi k_B T \tanh\left(\frac{b_z^I}{2 k_B T}\right)+\alpha \pi b_z^I}.
\end{align}
These results are reported in \cref{eq:Korringa,eq:T2} in the main text.
Since $s_{\pm}^\text{M}$ is complex, it causes oscillations in addition to the exponential decay, the spin's precession in the magnetic field.
Writing $s_{\pm} = -\frac{1}{T_2} + i \omega_\pm$, we capture this precession in the phase factors
\begin{align}
&\omega_\pm = \text{Im}\left[s_{\pm}^\text{M}\right]\notag\\
&= \mp \frac{2\alpha b_z^I}{\hbar}\left\{\ln\left(\frac{2\pi k_B T}{\xi'}\right)+\text{Re}\left[\psi\left(1\mp\frac{i b_z^I}{2\pi k_B T}\right)\right]\right\}.
\label{eq:phase}
\end{align}
Combining all the results leads to the Markovian decay of the reduced density matrix stated in \cref{eq:markovz,eq:markovpm}.
%

%
%

\section{Non-Markovian poles} \label{app:nonmarkovz}
Apart from the isolated Markov pole, we expect to find an array of poles close to the poles of the digamma function $\psi$. Using the series expansion for $\psi(1+z)$,
\begin{align} \label{eq:psi_expansion}
    \psi(1+z)=-\gamma+\sum_{n\geq1}\frac{z}{n(z+n)},
\end{align}
we see that the poles lie at negative integers $z=-n$. Again, we solve $s+iF_1(s)+iF_2(s)=0$ up to $O(\alpha)$. For $b_z^I=0$, the digamma function to investigate is $\psi(1+~s/2k_B T)$ and therefore we seek poles of the form $s_n=-2\pi k_B T n + p$, where $n \geq 1$ is an integer and $p$ is a small correction. Using the expansion of \cref{eq:psi_expansion}, we find to order $\alpha$ that $p$ is independent of $n$ and given by
\begin{align}
    p=-8\alpha \pi k_B T.
\end{align}
The residue corresponding to $s_n$ is, to order $\alpha$,
\begin{align}
    \text{Res}\left(s_n\right)=\frac{4\alpha}{n} e^{s_n t} = \frac{4\alpha}{n} e^{-2\pi k_B T(n+4\alpha)t}.
\end{align}
Here and for the remainder of this appendix we set the constant amplitudes $[\rho_z(t=0)-\rho_z^\text{eq}] = \rho_\pm(t=0) = 1$ and reintroduce them only in the final results.

The sum over all residues provides the non-Markovian time evolution.
Since at $b_z^I=0$ the evolution of $\rho_z(t)$ and $\rho_\pm(t)$ through \cref{eq:rhoz,eq:rhopm} is identical
we find that (reintroducing $\hbar$)
\begin{align}
    \rho^\text{nM}_{z}(t) &= \rho^{\text{nM}}_{\pm}(t)=\sum_{n\ge1}\frac{4\alpha}{n} e^{-2\pi k_B T(n+4\alpha)t/\hbar}\notag\\
    &=-4\alpha e^{-4\alpha t/\tau_T}\ln{\left(1-e^{-t/\tau_T}\right)},
\end{align}
where $\tau_T = \hbar / 2k_B T$ is the thermal time.

For $b_z^I>0$ we have to investigate in addition combinations with the shifted digamma functions $\psi(1+(s\pm i b_z^I)/2 k_B T)$. Each pole splits now into two and we must distinguish between the different $b_z^I$ dependences for $\rho_z$ and $\rho_\pm$. If we start with $\rho_z$, we write the pair of poles near $-2 \pi k_B T n$ as $s_{n,r}^z = - 2\pi k_B T n + p_r^z$, where $r = \pm$ and
\begin{equation}
	p_r^z = -4 \alpha \pi k_B T + r \sqrt{ (4 \alpha \pi k_B T)^2 - (b_z^I)^2}.
\end{equation}
The two corresponding residues become
\begin{align}
    \text{Res}\left(s_{n,r}^z\right)=
	\frac{2\alpha}{n} e^{s_{n,r}^z t} \left[1-r\frac{4\alpha\pi k_B T}{\sqrt{ (4 \alpha \pi k_B T)^2 - (b_z^I)^2}}\right].
\end{align}
The limit $b_z^I=0$ is correctly recovered since one of the residues then vanishes.
The summation over all $n$ provides then the non-Markovian evolution of $\rho_z$,
\begin{align}
	\rho_z^{\text{nM}}(t)
	&= \sum_{n\ge 1} \sum_{r=\pm}
	\frac{2\alpha}{n} e^{s_{n,r}^z t} \left[1-r\frac{4\alpha\pi k_B T}{\sqrt{ (4 \alpha \pi k_B T)^2 - (b_z^I)^2}}\right]
\notag\\
	&= -4\alpha e^{-4\alpha t/\tau_T}\ln{\left(1-e^{-t/\tau_T}\right)} h(t,b_z^I),
\end{align}
where we have defined
\begin{align}
	&h(t,b_z^I)
	=
	\cosh{\left(\frac{t}{\hbar}\sqrt{(4\alpha\pi k_B T)^2-(b_z^I)^2}\right)}\notag\\
    &-\frac{4\alpha\pi k_B T}{\sqrt{(4\alpha \pi k_B T)^2-(b_z^I)^2}}\sinh{\left(\frac{t}{\hbar}\sqrt{(4\alpha\pi k_B T)^2-(b_z^I)^2}\right)}.
	\label{eq:h_app}
\end{align}
%
\begin{figure}[t!]
    \includegraphics[width = \linewidth]{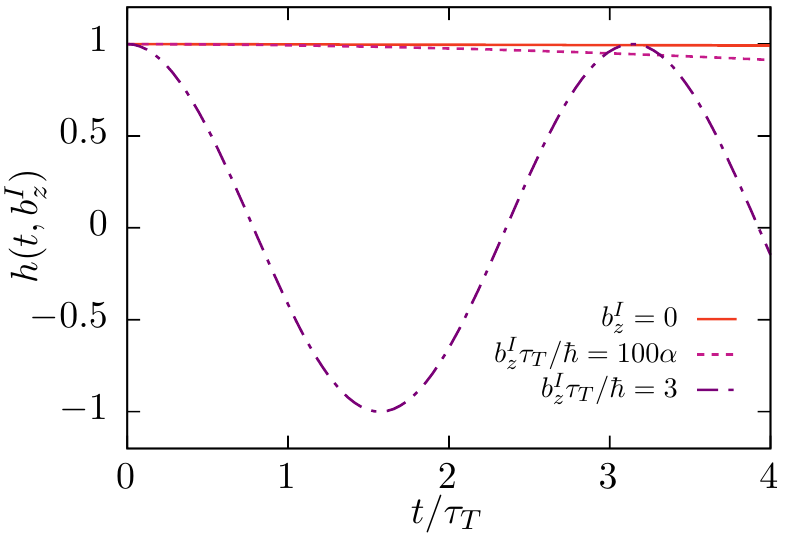}
    \caption{The function $h$ defined in \cref{eq:h_app} in the two limits of $b_z^I<\alpha\pi k_B T$ (purple, dashed) and $b_z^I>\alpha\pi k_B T$ (orange).}
    \label{fig:h}
\end{figure}
%
Reintroducing now the amplitude $[\rho_z(t=0)-\rho_z^\text{eq}]$, we obtain the result
\begin{align}
	\rho_z^{\text{nM}}(t)
	&= -4\alpha e^{-4\alpha t/\tau_T}\ln{\left(1-e^{-t/\tau_T}\right)} \notag\\
	&\ \ \ \times h(t,b_z^I) [\rho_z(t=0)-\rho_z^\text{eq}].
	\label{eq:nM_z_app}
\end{align}
For $\rho_\pm$, the only difference is that the arguments $s+ib_z^I$ and $s-ib_z^I$ of the digamma function terms are replaced by $s$ and $s\mp i b_z^I$.
This is exactly the same situation under the change of $b_z^I \to b_z^I/2$ and a shift of $s$ by $\mp i b_z^I/2$. Consequently, by exactly the same analysis,
\begin{align}
	\rho_\pm^{\text{nM}}(t)
	&= -4\alpha e^{\pm i b_z^I t/ 2\hbar} e^{-4\alpha t/\tau_T}\ln{\left(1-e^{-t/\tau_T}\right)} \notag \\
	&\ \ \ \times h(t,b_z^I/2) \rho_\pm(t=0).
	\label{eq:nM_pm_app}
\end{align}
The behavior of $h(t,b_z^I)$ on time scales relevant to the non-Markovian decay is shown in \cref{fig:h}.
For $b_z^I <~ 4\alpha \pi k_B T$, it contributes to the exponential decay but for larger magnetic fields it becomes oscillating with the frequency $\frac{1}{\hbar}{\sqrt{(b_z^I)^2-(4\alpha\pi k_BT)^2}}$, which at large fields or low $T$ saturates at the nuclear cyclotron frequency $b_z^I/\hbar$.


\bibliography{references}

\end{document}